\documentclass[aps,pra,twocolumn,superscriptaddress,eps2pdf]{revtex4-1}

\usepackage{epsfig}
\usepackage{amsfonts}
\usepackage{amssymb}
\usepackage{mathrsfs}
\usepackage{theorem}
\usepackage{amsmath}
\usepackage{times}
\usepackage{array,float}
\usepackage{longtable}

\newcolumntype{L}{>{\centering\arraybackslash}m{3cm}}

\usepackage{ifpdf}
\ifpdf
\usepackage{epstopdf}   
\fi

 \newtheorem{proof}{Proof}
 \newtheorem{theorem}{Theorem}
 \newtheorem{lem}{Lemma}
\newtheorem{corollary}{Corollary}
\newtheorem{exmp}{Example}[section]
\theoremstyle{plain}
\newtheorem{defin}{Definition}

 \newcommand{\ket}[1]{\ensuremath{\vert#1\rangle}}
 \newcommand{\bra}[1]{\ensuremath{\langle #1\vert}}

\newcommand{\kb}[2]{\ensuremath{\vert #1 \rangle \langle #2 \vert}}

\renewcommand{\vec}[1]{\ensuremath{\mathbf{#1}}}

\newcommand{\cliff}{\ensuremath{\sim_c}}
\newcommand{\pauli}{\ensuremath{\sim_{\mu}}}

\def\id{\mbox{\small 1} \!\! \mbox{1}}

\def\id{{\mathchoice {\rm 1\mskip-4mu l} {\rm 1\mskip-4mu l} {\rm 1\mskip-4.5mu l} {\rm 1\mskip-5mu l}}}

\begin{document}
\title{A unified framework for magic state distillation \\ and multiqubit gate-synthesis with reduced resource cost}

\author{Earl T.\ Campbell}
\email{earltcampbell@gmail.com}
\author{Mark Howard}
\affiliation{Department of Physics \& Astronomy, University of Sheffield, Sheffield, S3 7RH, United Kingdom.}

\begin{abstract}

The standard approach to fault-tolerant quantum computation is to store information in a quantum error correction code, such as the surface code, and process information using a strategy that can be summarized as distill-then-synthesize.  In the distill step, one performs several rounds of distillation to create high-fidelity logical qubits in a magic state.  Each such magic state provides one good $T$ gate.  In the synthesize step, one seeks the optimal decomposition of an algorithm into a sequence of many $T$ gates interleaved with Clifford gates.  This gate-synthesis problem is well understood for multiqubit gates that do not use any Hadamards. We present an in-depth analysis of a unified framework that realises one round of distillation and multiqubit gate synthesis in a single step. We call these synthillation protocols, and show they lead to a large reduction in resource overheads.  This is because synthillation can implement a general class of circuits using the same number of $T$-states as gate synthesis, yet with the benefit of quadratic error suppression. This general class includes all circuits primarily dominated by control-control-Z gates, such as adders and modular exponentiation routines used in Shor's algorithm.  Therefore, synthillation removes the need for a costly round of magic state distillation.  We also present several additional results on the multiqubit gate-synthesis problem.  We provide an efficient algorithm for synthesizing unitaries with the same worst-case resource scaling as optimal solutions.  For the special case of synthesizing controlled-unitaries, our techniques are not just efficient but exactly optimal.  We observe that the gate-synthesis cost, measured by $T$-count, is often strictly subadditive.  Numerous explicit applications of our techniques are also presented.
\end{abstract}

\maketitle

The topological surface code or toric code~\cite{dennis02} is the most widely known modern approach to quantum error correction.  Tolerating noise up to $1\%$~\cite{wang03,Rauss07}, it has established itself as the front-running proposal for quantum computation~\cite{Jones12,Fowler12,Nickerson14}.  However, it can not natively support fully universal quantum computation~\cite{Eastin09}.  Augmenting the surface code from a static device to a computer requires extra gadgets, which can be realised by a two-step process.  In the first step, magic state distillation is used to prepare encoded high-fidelity magic states~\cite{BraKit05}.  Each of these magic resources provides a fault-tolerant $T$-gate, also known as a $\pi/8$ phase gate.  In the second step, we decompose any desired unitary into a sequence of $T$-gates and Clifford gates, using gate-synthesis techniques to minimise the required number of $T$-gates.  We paraphrase this paradigm as distill-then-synthesize.  

%[expendable] Quantum computers are certainly feasible, but are they also affordable?  
After the initial discovery of Reed-Muller protocols for magic state distillation~\cite{Knill05,BraKit05},  recent years brought several innovations that reduced the cost of magic state distillation.   Next came the $10 \rightarrow 2$ protocol of Meier.~\textit{et al}~\cite{Meier13}, followed by the triorthgonal codes of Bravyi and Haah~\cite{Bravyi12}.  The Bravyi-Haah magic state distillation (BHMSD) protocol converts $3k+8$ magic states into $k$ magic states with quadratic error suppression, and will be our standard benchmark throughout.   Concatenating BHMSD two or three times, will suppress error rates from $10^{-4}$ to between $10^{-10}$ and $10^{-15}$, which suffices for many near term applications.  Once below very small error rates, multilevel distillation~\cite{jones13b} can further improve distillation yields, though it requires much larger circuits.  

\begin{figure*}
	\includegraphics{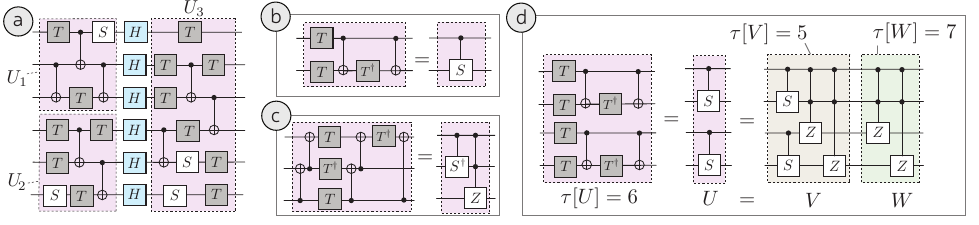}
	\caption{Example circuits.  (a) Complex circuits from Clifford+$T$ gate with subcircuits $\{ U_1, U_2, U_3 \}$  interspersed by Hadamard gates.  Subcircuits contain only control-NOT, $S$ and $T$ gates.  (b) Exact gate-synthesis of CS gate using 3 $T$-gates. (c) Exact gate-synthesis of the combined CS gate and CCZ gate using 4 $T$-gates. This  circuit is taken from Ref.~\cite{selinger13} and often referred to as $\mathrm{tof}^*$. (d) A pair of CS gates using 6 $T$-gates, and we illustrate $\mu[U]=5 < \tau[U]$ using its decomposition into $U=VW$  where $W$ contains only CCZ gates and $V$ attains $\tau[V]=5$.  These claims are proven later in Example~\ref{2NCS}.}
\label{fig:circuits}
\end{figure*}

Gate synthesis has undergone an even more impressive renaissance, making huge leaps forward since the early days of the Solovay-Kitaev theorem~\cite{kitaev02,dawson05}.  For synthesis of single qubit gates, optimal protocols have been found~\cite{kliuchnikov13,RS14,bocharov15}.   Here we are primarily interested in the multiqubit gate-synthesis problem~\cite{amy13,selinger13,amy14,maslov16,kliuchnikov13,bocharov15}.  For multiqubit circuits generated by CNOT and $T$ gates, optimal synthesis is well characterised~\cite{amy13,amy14,amy16,maslov16}, though no efficient solver exists for large circuits.   This multiqubit gate set requires Hadamards to acquire universality, and so gate-synthesis can be applied to subcircuits separated by Hadamards as shown in Fig.~(\ref{fig:circuits}a). This class of multiqubit gates is finite and can be exactly synthesized from the relevant gate set.  That is, there is no approximation error in this multiqubit synthesis problem and any noise arises from imperfections in the $T$ gates used.

However, the anticipated resource cost for fault-tolerant quantum computing remains formidable and we seek further reductions.  To date, most of this progress came about by treating magic state distillation and gate-synthesis as distinct puzzles.  However, one can circumvent the need for subsequent synthesis.  As an alternative to inexact synthesis of single qubit rotations, one can prepare special single qubit resources~\cite{duclos12,landahl13,duclos15,campbell16}.  In the multiqubit setting, the only known alternative approach prepares the resource state for a Toffoli gate~\cite{jones13b,eastin13,Paetznick13}.  This work inspired us to ask whether one can directly distill resources for a general class of multiqubit gates.

Here we present a general framework for implementing error-suppressed multiqubit circuits generated by CNOT and $T$ gates.  Our approach fuses notions of phase polynomials used in multiqubit gate synthesis~\cite{amy16} with a generalisation of Bravyi and Haah's triorthogonal $G$-matrices~\cite{Bravyi12}.  Our work reveals mathematical connections between these concepts, showing  our protocols to be formal unifications of previous of gate-synthesis and distillation protocols.  For single-qubit small-angle rotations, schemes like~\cite{landahl13,duclos15,campbell16} share some similarity with our current work, insofar as the need for subsequent synthesis is removed. The protocols in~\cite{jones13b,eastin13} are closer in spirit to our work as multi-qubit synthesis for the Toffoli (only) is implicitly performed, but our work makes the connections to synthesis both explicit and general.  On a practical level, synthillation is never more expensive than traditional distill-then-synthesize.  But, for a broad and important class of circuits, synthillation effectively eliminates the need for one round of distillation.  For many applications, we need only two or three rounds of BHMSD, so removing one round is a significant advance.  Asymptotically, one round of BHMSD uses three raw copies per output, and so by this metric our approach reduces overheads by approximately a third. We emphasise that this resource saving is benchmarked against optimal gate-synthesis, and so is cumulative with resource saving made over naive, suboptimal approaches to gate-synthesis.  The synthillation protocol is also compatible with module-checking~\cite{Ogorman16}, which offers further savings in some regimes.  We also present several techniques and efficient algorithms for finding gate-synthesis decompositions, which naturally feeds into our synthillation protocol.  In general, optimal gate-synthesis appears to be a hard problem, but we make progress by focusing on easy special classes and looking for near-optimal solutions.  

Our first section begins by formalising the exact multiqubit synthesis problem, and outlining our key results.  Sec.~\ref{Sec_framework} presents the synthillation protocol.  Sec.~\ref{secCC} provides the proofs for our gate-synthesis results.  Sec.~\ref{sec_Applications} goes into a detailed study of several concrete applications.  We close with Sec.~\ref{discuss}, discussing the broader context. All calculations and examples presented here can be reproduced using a Mathematica script in our supplementary material~\cite{CampSupp16}. A more concise account of the synthillation protocol is also available~\cite{Camp16b}. 

We remark that there are several ideas on how to circumvent magic state distillation~\cite{bombin:topoDistil,bombin:codeDeform,Paetznick13,Bombin13,bombin15,OConnor14}.  While these approaches save on the costs associated with magic state distillation, they all incur additional costs that are not immediately apparent.  For instance, typically these proposals require extra allocation of resources toward error correction.  So far, no alternative has been quantifiably shown to compete with two-dimensional topological codes combined with distill-then-synthesize.  In particular, no alternative has come close to the $1\%$ threshold of the surface code, with current numerics pointing toward 3D gauge colour codes possessing a threshold that is worse by an order of magnitude~\cite{brown15,bravyi15}.  This further motivates expanding the repertoire of techniques within the magic states paradigm.

\section{Overview}
\label{sec_Overview}

The magic states model was first formalized by Bravyi and Kitaev~\cite{BraKit05}.  It assumes certain operations are ideal,  free resources.  The model is justified because these operations are natively protected against noise in many error correcting codes, including the 2D topological codes such as the surface code and 2D colour codes.  The protected operations are called Clifford operations and  include: preparation of $\ket{0}$ states, measurement of Pauli-spin operators (elements of the Pauli group $\mathcal{P}$), unitaries in the Clifford group (denoted $\mathcal{C}$, the normalizer of the Pauli group), classical randomness and feed-forward. Stabilizer states can be reached from $\ket{0}$ states with Clifford unitaries and also constitute free resources.   In contrast, non-stabilizer states and non-Clifford unitaries are not natively protected, and so not free from noise and constitute costly resources. To obtain high-fidelity non-Clifford operations, such as the $T$-gate or preparation of magic $\ket{T}:=T\ket{+}$ states, requires several layers of magic state distillation, with each layer comprising many Clifford operations.  As such, the cost of magic states is significantly more than a Clifford operation. Throughout we measure resources by counting raw, noisy $\ket{T}$ states consumed.  This does not provide the full story as Clifford costs are not entirely negligible~\cite{fowler13,Ogorman16,maslov16}, but provides a good starting point for conceiving new protocols.  Throughout, we will often refer to a factor 3 saving in $T$-costs, and ask the reader to keep in mind that the full resource saving could be much greater than this.

We denote $\mathcal{C}^*$ for the subgroup of the Clifford group, which can be implemented with CNOTs and $S$ gates, where
\begin{equation}
	S = \left( \begin{array}{cc}
 1 & 0 \\
0 & i 	
 \end{array}
\right)	.
\end{equation}
We define the $T$ gate as
\begin{equation}
	T = \left( \begin{array}{cc}
 1 & 0 \\
	0 & \omega 	
 \end{array}
\right)	 ,
\end{equation}
with $\omega= \exp( i \pi / 4) $.  Composing gates in $\{ \mathcal{C}^*, T \}$, it was found~\cite{amy13} that all unitaries in the augmented group can be decomposed as  $V_{\mathrm{CNOT}} U_F$ where $V_{\mathrm{CNOT}}$ is some sequence of CNOT gates and $U_F$ belongs to a special class of diagonal unitaries.   We define this special class as $\mathcal{D}_3$, with gates in this group having the form 
\begin{equation}
    U_F = \sum_{\vec{x} \in \mathbb{Z}_2^k} \omega^{F(\vec{x})} \kb{ \vec{x} }{ \vec{x} },
\end{equation}
where $\ket{\vec{x}}$ is a computational basis state labelled by a binary string $\vec{x}^T=(x_1, x_2, \ldots , x_k)$, and $F$ is a cubic polynomial $F:\mathbb{Z}_2^k \rightarrow \mathbb{Z}_8$ of a particular form
\begin{align}
	F (\vec{x}) & =  L(\vec{x})  +  2 Q(\vec{x}) + 4 C(\vec{x}) \pmod{8}	,
\end{align}
where $L$, $Q$ and $C$ are linear, quadratic and cubic polynomials.  Explicitly, 
\begin{align}
\label{function_form}
	F (\vec{x}) & =  \sum_{i} l_i x_i  +  2 \sum_{i<j}  q_{i,j} x_i x_j 	\\ \nonumber
 & + 4  \sum_{i<j<k}  c_{i,j,k}  x_i x_j x_k \pmod{8}.
\end{align}
where the coefficients $l_i, q_{i,j}, c_{i,j,k}$ are integers defined modulo 8.  Sometimes we will refer to this as a weighted polynomial because the degree $m$ terms have coefficients that are weighted by $2^{m-1}$.  When it is clear from the context we drop the $F$ subscript from $U$, and at times it will be necessary to instead write $F_U$ as the function corresponding to $U$.  We will show later that the $U_F \in \mathcal{D}_3$  gates reside in the $3^{\mathrm{rd}}$ level of the Clifford hierarchy~\cite{CliffHier}, which explains our choice for the subscript 3.  We can directly infer that $U_F$ can be decomposed as $U_L  U_Q  U_C$ where $U_L$ contains only $T$ gates, $U_Q$ contains only control-$S$ gates (CS or short) and $U_C$ contains only control-control-Z gates (CCZ).  All these gates are diagonal in the computational basis with $U_{\mathrm{CS}}=\mathrm{diag}(1,1,1,i)$ and $U_{\mathrm{CCZ}}=\mathrm{diag}(1,1,1,1,1,1,1,-1)$.  We find a special role is played by unitaries composed of CCZ gates, and denote this subgroup as $\mathcal{D}_3^C$, where the superscript $C$ indicates that the associated weighted polynomial has only cubic terms, and so is a homogeneous cubic polynomial.  The gate set $\{\mathcal{C}^*, T\}$ is not universal, but becomes universal when $\mathcal{C}^*$ is promoted to the full Clifford group by including the Hadamard. The strategy of multiqubit gate synthesis is to take a universal circuit and partition it into subcircuits composed from $\{\mathcal{C}^*, T\}$ segmented by Hadamards, as illustrated in Fig.~(\ref{fig:circuits}a). From this one then optimises the decomposition of these subcircuits.

We define the $T$-count as following.
\begin{defin}
\label{tauDEF}
	For any $U \in \mathcal{D}_3$ we define the ancilla-free $T$-count as
\begin{equation}
	\tau[U]:= \min \{ t |   U=C_1 T_1 C_2 \ldots T_t C_n;     \{C_1, \ldots C_t \} \in \mathcal{C}^*  \}	.
\end{equation}
\end{defin}
It is possible to use fewer $T$-gates by exploiting ancilla.  Though, to the best of our knowledge, there is not yet a general toolbox for ancilla-assisted gate-synthesis and only a few such protocols are known (see e.g. Ref.~\cite{jones13b,paetznick14}).  In contrast, $\tau[U]$ is well understood and we have techniques for achieving optimality~\cite{amy16}.  We are interested solely in reducing $T$-counts, and do not consider $T$-depth or Clifford resources in our assessments of optimality. In Fig.~(\ref{fig:circuits}b) we show an optimal decomposition for realising a CS gate, and Fig.~(\ref{fig:circuits}c) shows an optimal decomposition for a combined CS$^\dagger$  gate and CCZ gate.  Individually, a CS$^\dagger$  gate require 3 $T$-gates and a CCZ gate requires 7 $T$-gates, but the composite circuit shown calls for only 4 $T$-gates where a naive composition of CS$^\dagger$ and CCZ would have used 10 $T$-gates.  The benefits of our synthillation protocol will be additional to such smart reductions in $T$-gates, and will use many of the same mathematical tools as gate-synthesis.

We find that CCZ gates are more amenable to resource savings than other $\mathcal{D}_3$ gates, and so introduce another measure of circuit complexity
\begin{defin}
\label{muDef}
	For any $U \in \mathcal{D}_3$ we define
\begin{equation}
	\mu[U]:= \min \{ \tau[V] | U=VW , W \in  \mathcal{D}^{C}_3, V \in  \mathcal{D}_3  \}	
\end{equation}
where $\mathcal{D}^{C}_3$ is the subgroup of $\mathcal{D}_3$ composed of CCZ gates.
\end{defin}
Clearly, $\mu[U] \leq \tau[U] $ since we can always set $W=\id$ and $V=U$.  Furthermore, if $U \in \mathcal{D}^{C}_3$ then $\mu[U]=0$ by setting $W=U$ and $V=\id$.  However, the $U=VW$ decomposition can be more counterintuitive.  In Fig.~(\ref{fig:circuits}d), we show a circuit where $U$ contains no CCZ gates, yet the minimisation to find $\mu[U]$ must use a decomposition where both $V$ and $W$ contain CCZ gates.  Having defined $\tau$ and $\mu$, we can state our main result
\begin{theorem}[The synthillation theorem]
\label{thm_prots}
Let $\{U_1, U_2, \ldots , U_l \}$ be a set of unitaries in the family $\mathcal{D}_3$, and $U = \otimes U_j$.  The synthillation protocol can implement $\{U_1, U_2,\ldots U_l \}$ with probability $1-n\epsilon+ O(\epsilon^2)$ and error rate $O(\epsilon^2)$ using 
\begin{equation}
    n = \tau[ U ] + 2 \mu[ U ] + \Delta \leq 3 \tau[U] + \Delta, 
\end{equation}
noisy $T$-states of initial error rate $\epsilon$, where $\Delta$ is a constant in the range $0 \leq \Delta \leq 11$.
\end{theorem}
  The constant $\Delta$ is bounded and so becomes unimportant in the limit of large circuits.  The synthillated $U_j$ need not be implemented in parallel, each unitary $U_j$ maybe injected into a circuit at any point.  See Fig.~\ref{fig:circuits}a for an example set $\{ U_1, U_2, U_3 \}$ that are not injected as a tensor product, though the synthillation cost is determined by $U=U_1 \otimes U_2 \otimes U_3$.  It is important to recognise that $\epsilon$ is error rate on the magic states used rather than a measure of synthesis precision.  For inexact synthesis problems, $\epsilon$ is often used to quantify the precision of an implemented unitary relative to a target unitary.  In this context, synthesis is exact.  Both our protocol and gate-synthesis~\cite{amy13,selinger13,amy14,maslov16,kliuchnikov13,bocharov15} will implement a perfect $U$ when supplied with perfect magic states. Given imperfect magic states with error $\epsilon$, synthillation realises $U$ with quadratically suppressed $O(\epsilon^2)$ error, whereas using the same magic states gate-synthesis would lead to a $O(\epsilon)$ implementation of $U$.  

\begin{figure}
	\includegraphics{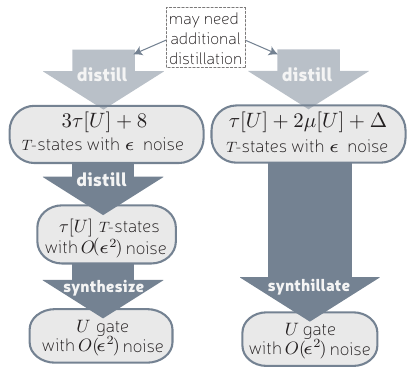}
	\caption{Overview of the comparison between: (left) conventional distill-then-synthesize, using BHMSD and optimal gate-synthesis and (right) synthillation.  We study the final step of error-correction, but approaches may need additional precursor rounds of distillation to reach target fidelity.  Typically, both approaches need an equal number of precursor rounds.}
\label{synth_flow}
\end{figure}

Therefore, we instead compare synthillation against distill-then-synthesize, which is one round of distillation followed by gate-synthesis.  Now both approaches yield $O(\epsilon^2)$ error, but have different resource overheads and are summarised in Fig.~\ref{synth_flow}.  Asymptotically, our approach is never more expensive than using a round of BHMSD followed by gate synthesis, which would cost $n = 3 \tau[ U ]$ ignoring additive constants.      Whereas, if $\mu[ U ] \ll \tau[ U ]$ synthillation costs $\sim 1/3$ the price of using BHMSD with gate-synthesis.  This maximum saving is attained whenever $U \in \mathcal{D}_3^{C}$ as then $n=\tau[ U ]+\Delta$.  This class of circuits is common as quantum algorithms often contain components that consist of classical reversible logic achieved using only Toffoli gates, CNOT gates and NOT gates.  For instance, modular exponentiation is simply classical logic and also amounts to the dominant resource cost in Shor's algorithm~\cite{kitaev02,fowler04}.  Furthermore, Toffoli and Hadamard form a universal gate set, so the gate set $\{ \mathcal{D}_3^{C}, H \}$ is universal. Beyond Toffoli  circuits, there are many other cases where we obtain close to this $1/3$ saving, which is ensured by the following  
\begin{theorem}
\label{circuits}
	For all  $ U \in \mathcal{D}_3$ acting on $k$ qubits, we have $\mu[U] \leq k+1$.  Furthermore, there exists a $\mathrm{poly}(k)$ algorithm for finding both a $U=VW$ decomposition (with $\tau[V]=\mu[U]$ and $W\in \mathcal{D}_3^C$) and also an optimal synthesis of $V$ using Clifford+$T$ gates.
\end{theorem}
We see this theorem at work in Fig.~(\ref{fig:circuits}d), where a 4 qubit circuit has $\mu[U]=5<\tau[U]$ even though $U$ does not contain any CCZ gates.  More generally, this shows that $\mu$ scales at most linearly with the number of qubits, whereas Amy and Mosca~\cite{amy16} showed that $\tau$ scales at most quadratically.  This quadratic scaling tells us that complex circuits may have $k \ll \tau[U]$ which entails $\mu[U] \ll  \tau[U]$.  In such cases, the distillation cost becomes comparable to the gate synthesis cost.  Our proof of Thm.~\ref{circuits} reduces it to a matrix factorization problem, which can be solved using a known algorithm.  This is remarkable because the optimisation problem for $\tau$ is believed to be a hard problem, see Ref.~\cite{amy16} and Sec~\ref{PP}.  We prove Thm.~\ref{circuits} in Sec.~\ref{Bmatrixfast}.  

Since finding the optimal $\tau$ is difficult, we need efficient algorithms for near-optimal decompositions.  We will show that a fast algorithm exists giving approximation solutions
\begin{theorem}
\label{subOPT}
	Let  $ U \in \mathcal{D}_3$ acting on $k$ qubits. There exists a $\mathrm{poly}(k)$ algorithm that finds a decomposition of $U$ in terms of Clifford+$T$ gates, with $\tau_{\mathrm{fast}}[U]$ uses of $T$-gates where 
\begin{equation}
	\tau[U] \leq \tau_{\mathrm{fast}}[U] \leq \frac{k^2}{2}+\frac{k}{5}-11.
\end{equation}
\end{theorem}
Previous efficient algorithms do not have such scaling.  For instance, $T_{PAR}$~\cite{amy14} has no proven upper bound in $T$-count, though in practice may perform well.  Implicit in Ref.~\cite{amy16} is an efficient algorithm with a maximum $O(k^3)$ cost, but this still leaves a significant gap compared to the scaling of optimal solutions.  

While it is believed that in general the optimal gate synthesis problem is hard, special cases can be tractable.  In Sec.~\ref{optimal_cont_U} we consider controlled-unitaries in $\mathcal{D}_3$ and show this subclass can be solved efficiently and optimally, with $\tau$ upper bounded by $2k+1$ for $k$ qubit unitaries. 

We also observe that $\tau$ does not behave additively, so there are unitaries $U_1$ and $U_2$ such that $\tau[U_1 \otimes U_2] < \tau[U_1]+\tau [U_2]$.  While it is clear that composed gates $U_1 U_2$ can be subadditive in cost, it seems remarkable that entirely disjoint circuits enjoy a reduction in resource costs.  This subadditivity is reminiscent of similar phenomena seen in different resource theoretic settings.

In the final section we tackle concrete applications. Previous results show $O(\epsilon^2)$ error Toffoli gates are possible using 8 $T$ states.  We find error suppressed Toffoli gates are available at an asymptotic cost of 6 $T$-states each, which is partly due to aforementioned subadditivity.  As a mainly pedagogical exercise we consider many control-S gates.  Last we consider a family of circuits composed of CCZ gates, where optimal gate-synthesis offers a saving of naive gate-synthesis, and we obtain a further factor 3 reduction in resource by using synthillation.

\section{The general framework}
\label{Sec_framework}

\subsection{The Clifford hierarchy and Clifford equivalence}
\label{sec_hierarchy}

Here we review the Clifford hierarchy, introduce an equivalence relation and fix some notation.  The $j^{\mathrm{th}}$ level of the Clifford hierarchy is defined as
\begin{equation}
	\mathcal{C}_j := \{ U | U^\dagger P U P^\dagger  \in \mathcal{C}_{j-1}  ; \forall P \in \mathcal{P} \}	,
\end{equation}
where $\mathcal{P}$ is the Pauli group and we terminate the recursion with $\mathcal{P}=\mathcal{C}_1$.  The familiar Clifford group is $\mathcal{C}_2$.  For higher levels of the hierarchy we get non-Clifford gates.  Here we concern ourselves with non-Cliffords from the third of the hierarchy.  Specifically, we have defined the group $\mathcal{D}_3 $, which is readily verified to be the diagonal subgroup of $\mathcal{C}_3$.  Furthermore, we have that for all $U \in \mathcal{D}_3$, the gate $U^2$ is in the diagonal Clifford group.  In terms of weighted monomials we have $U_F^2 = U_{2F}$.  We give further details in App.~\ref{App_CliffHier}.  The Clifford hierarchy is important as it has been shown that gates in $\mathcal{C}_3$ can be performed by teleportation using Clifford operations and a particular resource state~\cite{Zhou00}.  When the gate is also diagonal this resource is simply $U \ket{+}^{\otimes k}$.

We say two unitaries $U$ and $V$ are Clifford equivalent whenever there exist Cliffords $C$ and $C'$ such that $U=C V C'$.  Since $U_{2 \tilde{F}}$ is a Clifford for any weighted polynomial $\tilde{F}$, we know that $U_{F}$ and $U_{F} U_{2 \tilde{F}}=U_{F+2 \tilde{F}}$ are Clifford equivalent.  In other words, two unitaries $U_F$ and $U_{F'}$ are Clifford equivalent whenever there exists an $\tilde{F}$ such that $F=F' + 2 \tilde{F} \pmod{8}$.  In such cases we write $F \cliff F'$ where $\cliff$ is an equivalence relation.  It follows immediately that if $F \cliff F'$ then $\tau[U_F]=\tau[U_{F'}]$, and we reiterate that $\tau$ was specified in Def.~\ref{tauDEF}.  Since $\tau$ and $\cliff$ are closely related, it is natural to ask whether $\mu$ (recall Def.~\ref{muDef}) is related to some equivalence relation?  In Sec.~\ref{Bmatrixfast} we introduce such an equivalence relation. Lastly, we use $\mathrm{col}(M)$ to denote the number of columns in matrix $M$ and $\mathrm{row}(M)$ to denote the number of rows in matrix $M$.

 \subsection{Quantum codes, encoders and quasitransversality}

Central to synthillation are quantum codes with a special property we call quasitransversality.  Here we define a quantum code in the $G$-matrix formalism, generalising the work of Bravyi and Haah~\cite{Bravyi12}.  To specify a code we use a binary matrix $G$ partitioned into $K$ and $S$. 
\begin{defin}
\label{Gcode}
Let $G$ be a binary matrix that is full $\mathbb{Z}_2$-rank with $n$ columns and $k+s$ rows that is partitioned into $K$ and $S$ so that $G=(\frac{K}{S})$.  We define a quantum code with logical basis states
\begin{equation}
    \ket{\vec{x}_L} := \frac{1}{2^{s/2}}  \sum_{\vec{y} \in \{0,1 \}^{s}}  \ket{ K^T \vec{x} \oplus S^T \vec{y} },
\end{equation}
This is an $[[n,k,d]]$ code where $n$ is the number of columns in $G$, $k$ is the number of rows in $K$, and with some distance $d$. 
\end{defin}
We note that the $j^{\mathrm{th}}$ element of $K^T \vec{x} \oplus S^T \vec{y}$ is explicitly 
\begin{equation}
	 \left( K^T \vec{x} \oplus S^T \vec{y} \right)_j= \sum_{i=1}^{k}K_{i,j}x_i+\sum_{i=1}^{s}S_{i,j}y_i \pmod 2 .
\end{equation}
We say the code is trivial if the $S$ partition is empty, which entails $d=1$.  Bravyi and Haah considered binary matrices split according to row weight, with odd weight rows in $K$ and even weight rows in $S$.  We do not make this assumption, but will later impose a more complex condition dependent on the desired unitary.

Next, we review properties of encoder circuits used to prepare these quantum codes states. We  use that for any invertible binary matrix $J$, there exists~\cite{Dehaene03,patel03,maslov07} a CNOT circuit $E_J$ such that 
\begin{equation}
	E_J = \sum_{\vec{z}} \ket{J^T \vec{z}} \bra{\vec{z}},
\end{equation}
In addition to its action on the computational basis, we track how these unitaries alter Pauli-$Z$ operators. To describe $Z$ operators acting on many qubits we use $Z[\vec{e}]:=\otimes_{j=1}^n Z_j^{e_j}$ where $\vec{e}$ is some binary vector.  Therefore, 
\begin{equation}
		Z[\vec{e}] = \sum_{\vec{v} \in \mathbb{Z}_2^n} (-1)^{\langle \vec{v}, \vec{e} \rangle}\kb{\vec{v}}{\vec{v}},
\end{equation}
where throughout $\langle \ldots , \ldots \rangle$ is the inner product satisfying  $\langle \vec{v}, \vec{e} \rangle = \sum_{j} v_j e_j \pmod{2}$. The Clifford $E_J$ affects the conjugation
\begin{align}
		E_J^\dagger Z[\vec{e}]	E_J & = \sum_{\vec{u},\vec{v}} \kb{\vec{u}}{J^T\vec{u}} Z[\vec{e}] \kb{J^T\vec{v}}{\vec{v}} \\ \nonumber
 & = \sum_{\vec{u},\vec{v}} (-1)^{\langle J^T\vec{v},\vec{e} \rangle} \kb{\vec{u}}{J^T\vec{u}} J^T\vec{v} \rangle \bra{\vec{v}} \\ \nonumber
 & = \sum_{\vec{v}}  (-1)^{\langle J^T\vec{v},\vec{e} \rangle} \kb{\vec{v}}{\vec{v}} . 
\end{align}
We use that the inner product satisfies $\langle J^T\vec{v},\vec{e} \rangle = \langle \vec{v},J\vec{e} \rangle $ to conclude that
\begin{equation}
		E_J^\dagger Z[\vec{e}]	E_J = Z[J\vec{e}].
\end{equation}
For a quantum code, the matrix $G$ will not be square, and so cannot be invertible.  However, there will always exist an invertible $J$ that completes $G$, so that
\begin{equation}
	J = \left( \begin{array}{c}
 	G \\
	M
 \end{array}
 \right) = \left( \begin{array}{c}
 	K \\
 	S \\
	M
 \end{array}
 \right)	,
\end{equation}
for some $M$. We consider $J$ to act on a partitioned bit string composed of $\vec{x}$, $\vec{y}$, and $\vec{z}$, so that
\begin{equation}
	J^T \left( \begin{array}{c}
\vec{x} \\ 
\vec{y} \\ 
\vec{z} \\ 
 \end{array}
\right) = K^T\vec{x} \oplus S^T \vec{y}	 \oplus M^T \vec{z}
\end{equation}
and
\begin{equation}
		E_J \ket{ \vec{x} }\ket{\vec{y}}\ket{\vec{z}} =\ket{K^T \vec{x} \oplus S^T \vec{y} \oplus M^T \vec{z}},
\end{equation}
and similarly
\begin{align}
\label{encoder_pauli_conj}
	   E_J^\dagger Z[ \vec{e}]	E_J &  = Z[ K\vec{e} ] \otimes Z[ S\vec{e}]\otimes  Z [M\vec{e} ].
\end{align}
For the special case $\vec{z}=\vec{0}$, we have $M^T \vec{0} = \vec{0}$ and so
\begin{align}
	 E_J \ket{ \vec{x} }\ket{ \vec{y} } \ket{0}^{\otimes n-k-s}	 & = E_J \ket{\vec{x}}\ket{\vec{y}} \ket{\vec{0}}\\ \nonumber
		& =  \ket{ K^T\vec{x} \oplus S^T\vec{y} } 
\end{align}
Therefore, with appropriate ancilla qubits set to $\ket{0}$, all completions of $G$ behave identically, independent of the choice of $M$.  From here onwards, we use $E_G$ to denote any unitary with the above action.  We will often refer to $E_G$ as an encoder for the quantum code associated with $G$ because of the following
\begin{align} \nonumber
	E_G \ket{\vec{x}}	\ket{+}^{\otimes s}  \ket{0}^{\otimes  n-k-s}
& = \frac{1}{2^s} \sum_{\vec{y} \in  \mathbb{Z}_2^s} E_G \ket{\vec{x}}	\ket{\vec{y}} \ket{0}^{\otimes  n-k-s} \\ \nonumber
& = \frac{1}{2^s} \sum_{\vec{y} \in  \mathbb{Z}_2^s}  \ket{ K^T \vec{x} \oplus S^T \vec{y}  } \\ 
& = \ket{\vec{x}_L} .
\end{align}
This shows how logical stabilizer states can be prepared using unencoded stabilizer states and CNOT gates.  

Crucially important are quantum codes with the following property.
\begin{defin}
\label{natural}
Let $F$ be a weighted polynomial and $U_F \in \mathcal{D}_3$ the associated unitary. We say a quantum code is $F$-quasitransversal if there exists a Clifford $C$ such that $C T^{\otimes n}$ acting on the code realises a logical $U_F$.
\end{defin}
Transversal logical gates can be realised with product unitaries.  Here only the non-Clifford part is required to have product form, and the Clifford gate can be non-product, so we say they are quasitransversal. A sufficient condition for $F$-quasitransversality is the following.
\begin{lem}
\label{quasi_transversal}
Let $F$ be a weighted polynomial with associated $U_F \in \mathcal{D}_3$.  Let $G$ be a $(k+s)$-by-$n$ full $\mathbb{Z}_2$-rank matrix partitioned into $K$ and $S$.  The associated quantum code is $F$-quasitransveral if
\begin{equation}
\label{quasiTransCond}
		|  K^T \vec{x} \oplus S^T \vec{y} | \cliff F(\vec{x})  \pmod{8} .
\end{equation}
\end{lem}
Here we use $| \ldots |$ to denote the weight of a vector, so $|\vec{e}|:=\sum_j e_j$.  Before proving the lemma, let us unpack the notation.  The equation is evaluated $\pmod{8}$, but $K^T \vec{x} \oplus S^T \vec{y} $ is always evaluated $\pmod{2}$. Furthermore, this compact notation can be expanded out as
\begin{equation}
|  K^T \vec{x} \oplus S^T \vec{y} |   := \sum_{j} \left[ \sum_{i=1}^{k} K_{i,j}x_i +  \sum_{i=1}^{s} S_{i,j}y_j \pmod{2} \right] 
\end{equation}
Applying $T^{\otimes n}$ to an encoded state gives 
\begin{align}
	T^{\otimes n}\ket{\vec{x}_L} & = \frac{1}{2^{s/2}}  \sum_{ \vec{y} \in \mathbb{Z}_2^s} \omega^{|K^T \vec{x} \oplus S^T \vec{y}|} \ket{K^T \vec{x} \oplus S^T \vec{y} }.
\end{align}
Any diagonal Clifford $\tilde{C}$ acts as
\begin{equation}
	\tilde{C} \ket{\vec{x}}\ket{\vec{y}}\ket{\vec{0}}^{\otimes n-k-s}=\omega^{2\tilde{F}(\vec{x},\vec{y})}\ket{\vec{x}}\ket{\vec{y}}\ket{\vec{0}}^{\otimes n-k-s}
\end{equation}
for some $\tilde{F}$, where we set some qubits zero.  We define another diagonal Clifford $C:=E_G \tilde{C} E_G^\dagger$ so that 
\begin{align}
	C \ket{K^T \vec{x} \oplus S^T \vec{y}} &= E_G \tilde{C} E_G^\dagger \ket{K^T \vec{x} \oplus S^T \vec{y}} \\ \nonumber
&= E_G \tilde{C} \ket{\vec{x}}\ket{\vec{y}}\ket{\vec{0}} \\ \nonumber
&= \omega^{2\tilde{F}(\vec{x},\vec{y})} E_G \ket{\vec{x}}\ket{\vec{y}}\ket{\vec{0}}\\ \nonumber
&= \omega^{2\tilde{F}(\vec{x},\vec{y})}  \ket{K^T \vec{x} \oplus S^T \vec{y}} 
\end{align}
Therefore, the combined unitary $C T^{\otimes n}$ acts as
\begin{align}
C T^{\otimes n} \ket{\vec{x}_L}&
= \frac{1}{2^{s/2}} \sum_{\vec{y}}\omega^{|K^T \vec{x} \oplus S^T \vec{y}|+2\tilde{F}(\vec{x},\vec{y})}\ket{K^T \vec{x} \oplus S^T \vec{y}} .
\end{align}
The lemma assumes that  $|K^T \vec{x} \oplus S^T \vec{y}| \cliff F(\vec{x}) \pmod 8$, which is equivalent to the existence of an $\tilde{F}$ such that 
\begin{equation}
		|K^T \vec{x} \oplus S^T \vec{y}| + 2\tilde{F}(\vec{x},\vec{y}) = F(\vec{x}) \pmod 8.
\end{equation}
Furthermore, since $\omega^8 = 1$, the exponent of $\omega$ is can be taken modulo 8, and so 
\begin{equation}
		\omega^{|K^T \vec{x} \oplus S^T \vec{y}| + 2\tilde{F}(\vec{x},\vec{y})} = \omega^{F(\vec{x})} .
\end{equation}
Using this $\tilde{F}$ to specify $\tilde{C}$ and thereby $C$, we have 
\begin{align}
C T^{\otimes n} \ket{\vec{x}_L}&
= \frac{1}{2^{s/2}} \sum_{\vec{y}}\omega^{F(\vec{x})}\ket{K^T \vec{x} \oplus S^T \vec{y}} .
\end{align}
Since the phase no longer depends on $\vec{y}$, the phase can come outside the summation
\begin{align}
C T^{\otimes n} \ket{\vec{x}_L}
& = \frac{\omega^{F(\vec{x})}}{2^{s/2}}  \sum_{\vec{y}}\ket{K^T \vec{x} \oplus S^T \vec{y}} \nonumber \\
& =\omega^{F(\vec{x})} \ket{\vec{x}_L}.
\end{align}
This proves $F$ quasitransversality follows from the condition stated in the lemma.

\subsection{The synthillation protocol}
\label{synth}

\begin{figure*}
	\includegraphics{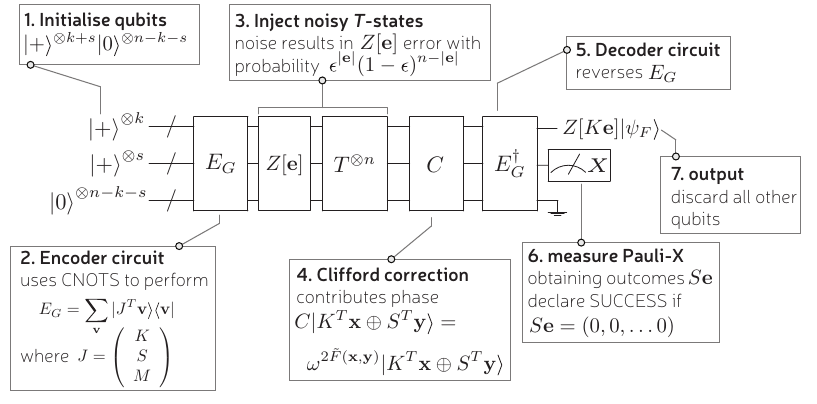}	
\caption{The main segment of the synthillation protocol, which prepares an error suppressed $\ket{\psi_F}$.  We follow this by using $\ket{\psi_F}$ to inject the correspond gate $U_F$ into a quantum algorithm.}
\label{Fig_prot}
\end{figure*}

Given a quasitransversal quantum code, we can construct protocols for preparing $U_{F}\ket{+}^{\otimes k}$ magic states.
\begin{theorem}
\label{thm_main}
Let $G$ be a $(k+s)$-by-$n$ full $\mathbb{Z}_2$-rank matrix.  Let $| K^T \vec{x} \oplus S^T \vec{y} |  \cliff F(\vec{x})  \pmod{8}$ so that the associated quantum code is $F$ quasitransversal. There exists a distillation protocol using only Clifford operations and $n$ noisy $T$-states with error rate $\epsilon$.  The protocol outputs the magic state $\ket{\psi_{F}} = U_{F}\ket{+}^{\otimes k}$ with error rate 
$O(\epsilon^d)$ where $d$ is the distance of the quantum code associated with $G$.   If $d>1$, then the success probability is $p_{\mathrm{suc}}=1 - n \epsilon + O(\epsilon^2)$.
\end{theorem}
The above is a key finding of this work, and essentially delegates the task of finding synthillation protocols to finding $G$ matrices with the required properties.  One can express Bravyi-Haah's notion of triorthogonality as  
\begin{equation}
	| K^T \vec{x} \oplus S^T \vec{y} |  \cliff \sum_{i=1}^k x_i	,
\end{equation}
and so our concept of quasitransversality is a generalisation thereof.  We discuss this point further in App.~\ref{App::TriOrtho}.

We describe the protocol as a quantum circuit in Fig.~\ref{Fig_prot}. We first show why the protocol works in the absence of noise.  Up to step 5 we have, 
\begin{align}
\label{ideal}
	 & E_G^\dagger  C T^{\otimes n} E_G  \ket{+}^{\otimes k+s}\ket{0}^{\otimes n} \\ \nonumber &= (U_{F}\ket{+}^{\otimes k}) \ket{+}^{\otimes s} \ket{0}^{\otimes n-k-s} ,
\end{align}
which follows directly from $F$ quasitransversality.    Without noise, the measured qubits  are in the $\ket{+}$ state and so yield ``+1" outcomes in step 6.  After discarding qubits in step 7 we are left with $\ket{\psi_F}=U_F \ket{+}^{\otimes k}$.

Now we consider noise.  The noisy $T$-gates can, by twirling, be ensured to only suffer from Pauli $Z$-noise.   Therefore, at step 3 we must add the operator $Z[\vec{e}]$ with probability $p(\vec{e})=\epsilon^{| \vec{e} |}(1-\epsilon)^{n-| \vec{e} |}$.  Recalling Eq.~(\ref{encoder_pauli_conj}) and using that $C$ commutes with $Z[\vec{e}]$ we have
\begin{align}
	E_G^\dagger C Z[\vec{e}] T^{\otimes n} E_G & = E_G^\dagger Z[\vec{e}]E_G  E_G^\dagger C  T^{\otimes n} E_G	 \\ \nonumber
&= Z[J\vec{e}]  E_G^\dagger C  T^{\otimes n} E_G	 ,
\end{align}
where $Z[J\vec{e}]=Z[K \vec{e}] \otimes  Z[S \vec{e}] \otimes Z[M \vec{e}] $.  Therefore, the noisy output differs by $Z[J\vec{e}]$ from the ideal case (see Eq.~\ref{ideal}) so that
\begin{align}
	 &  E_G^\dagger C Z[\vec{e}] T^{\otimes n} E_G  \ket{+}^{\otimes k+s}\ket{0}^{\otimes n} \\ \nonumber 
&=
		Z[J\vec{e}]( U_F\ket{+}^{\otimes k} \ket{+}^{\otimes s} \ket{0}^{\otimes n-k-s} \\ \nonumber 
&=
		(Z[K\vec{e}] U_F\ket{+}^{\otimes k}) (Z[S\vec{e}]\ket{+}^{\otimes s})(Z[M\vec{e}] \ket{0}^{\otimes n-k-s}) \\ \nonumber
 &=
		(Z[K\vec{e}] \ket{ \psi_F})( Z[S\vec{e}]\ket{+}^{\otimes s}) \ket{0}^{\otimes n-k-s}
\end{align}
where between the second and last line we have used $Z\ket{0}=\ket{0}$ to eliminate $Z[ M \vec{e}]$.  Some Pauli operators $Z[ S^T \vec{e}]$ will act nontrivially on the $\ket{+}^{\otimes s}$ qubits, flagging up the error. In step 6, we measure the qubits in the state $Z[S \vec{e}] \ket{+}^{\otimes s}$ obtaining the SUCCESS outcome only if $Z[S \vec{e}]=\id_s$ and so $S \vec{e}=(0,0,\ldots0)$.  Therefore, the success probability is
\begin{equation}
	p_{\mathrm{suc}} = \sum_{\vec{e} ,  S \vec{e}=(0,\ldots 0)} \epsilon^{|\vec{e}|}(1-\epsilon)^{n-|\vec{e}|}.
\end{equation}
The output state is $Z[K \vec{e}]\ket{ \psi_F}$ which is the correct state whenever $Z[K \vec{e}]=\id_k$ and so $K \vec{e}=(0,0,\ldots )$.  Therefore, the normalised error rate is
\begin{equation}
	\epsilon_{\mathrm{out}} =1-\frac{1}{p_{\mathrm{suc}}} \sum_{\vec{e} ,  K\vec{e}=(0,\ldots 0)} \epsilon^{|\vec{e}|}(1-\epsilon)^{n-|\vec{e}|}.
\end{equation}
For a distance $d$ code, we have that if $S \vec{e}=(0,0,\ldots , 0)$ and $K \vec{e} \neq (0,0,\ldots,0 )$ then $|\vec{e}| \geq d$. This allows us to conclude the scaling $\epsilon_{\mathrm{out}} = O(\epsilon^d)  $, and completes our proof of Thm.~(\ref{thm_main}). 

Given a matrix $G$, the above expressions allow us to find the exact expressions for $p_{\mathrm{suc}}$ and $\epsilon_{\mathrm{out}}$ by summing over all $\vec{e}$ meeting the criteria.  Typically, this sum will involve many terms and for large matrices could be computationally challenging.  However, the sums can be reduced to far fewer terms by using the MacWilliams identities to move to a dual picture.  This use of MacWilliams identities is a standard trick used within the field~\cite{BraKit05,Bravyi12,campbell12} and entails
\begin{equation}
\label{MacWillpsuc}
	p_{\mathrm{suc}} =\frac{1}{2^s} \sum_{\vec{e} \in \mathrm{span}(S)} (1-2\epsilon)^{|\vec{e}|},
\end{equation}
where we sum over all bits strings in the vector space generated by the rows of $S$, which we denote as $\mathrm{span}(S)$. 

\subsection{Constructing the codes}
\label{Sec_construct}

Thm.~\ref{thm_main} showed how to perform synthillation  given $G$ matrices satisfying certain conditions depending on the target unitary $U_F$.  The next step in our proof is to construct such $G$ matrices from submatrices.  We begin by introducing the building blocks.
\begin{defin}
	We say a binary matrix $A$ is a gate-synthesis matrix for unitary $U$ if
\begin{equation}
	| A^T \vec{x} | \cliff F(\vec{x})	
\end{equation}
where $F$ is the weighted polynomial for $U$.
\end{defin}
This definition is a simpler version of the quasitransversality condition of Eq.~(\ref{quasiTransCond}), because a gate-synthesis matrix does not have the additional degrees of freedom needed to suppress errors.  Throughout, we use $A$ to denote a gate synthesis matrix for $U$, so $| A^T \vec{x} | \cliff F_U(\vec{x})$.  Recall that Thm.~\ref{thm_main} and Def.~\ref{muDef} made use of a decomposition $U=VW$ where $W \in \mathcal{D}_3^C$, and so we use $B$ to denote the gate synthesis matrix for any such $V$, so that $| B^T \vec{x} | \cliff F_{V}(\vec{x})$. We find later that optimal matrices have columns numbering $\mathrm{col}(A)=\tau[U]$ and $\mathrm{col}(B)=\mu[U]$.  The next section discusses techniques for constructing $A$ and $B$, and to what extent optimal constructions can be found by an efficient algorithm.  However, for the purposes of this section, these matrices need not be optimal.  If suboptimal matrices are used, the resource cost is $n=\mathrm{col}(A)+ 2\mathrm{col}(B)+\Delta$.

The construction of distillation matrix $G$ and the value of the constant $\Delta$ vary depending on numerous features, leading to 11 different cases presented in Table~\ref{TABallCASES}.  Here we give an explicit proof of the result for three cases of increasing complexity. The remaining cases follow the same methodology with only minor changes. Before we begin the proofs, we review some of the basic tools.  For any weighted polynomial of the form $F(\vec{x}) = L(\vec{x}) + 2Q(\vec{x}) + 4C(\vec{x})$, we have that
\begin{enumerate}
	\item $2  F(\vec{x}) \cliff 0$; 
    \item  $2 F(\vec{x}) =  2L(\vec{x}) + 4Q(\vec{x}) \pmod{8} $ and so for homogeneous cubic functions $2F(\vec{x}) = 0 $;
    \item $4 F(\vec{x}) = 4L(\vec{x}) \pmod{8}$ and so for functions without a linear component   $4 F(\vec{x}) = 0 $;
    \item if $F_1 \cliff 0$  and $F_2 \cliff 0 $, then $F_1 F_2 \cliff 0$;
\end{enumerate}
Property 1 follows directly from the discussion in Sec.~\ref{sec_hierarchy}. Property 2 and 3 follows directly due to modulo 8 arithmetic.   The last property is also proven by similar expansions and degree counting. We shall also make use of the modular identity $\vec{u} \oplus \vec{v} = \vec{u} + \vec{v} - 2 \vec{u} \wedge \vec{v}$ where $\wedge$ is the element-wise product of two vectors. 

We begin by considering the simple case 9
\[ 
G_9 =  \left( \begin{array}{c}
K_9   \\ \hline
S_9  \end{array} \right)= \left( \begin{array}{c}
A   \\ \hline
\rule[-.3\baselineskip]{0pt}{10pt}  \vec{1}^T  \end{array} \right),
\]
where throughout $\vec{1}^T=(1,1,\ldots, 1)$ and the vector length should be clear from the context.  We remind the reader that bold font symbols are used for column vectors, and so row vectors carry a transpose. We have
\[
	|K_9^T\vec{x} \oplus S^T_9 \vec{y} | =|(A^T\vec{x}) \oplus ( y_1  \textbf{1} )|.
\] 
Using the modular identity, we have
\begin{align} \nonumber
	|(A^T\vec{x}) \oplus ( y_1  \textbf{1} )| = & |A^T\vec{x}| + | y_1  \textbf{1} |  \\ \label{prevanish}
 & - 2 | (A^T \vec{x}) \wedge ( y_1  \textbf{1} ) |  .
\end{align}
We notice that $| y_1  \textbf{1} | =  y_1  |  \textbf{1} |$, where $|  \textbf{1} | = \mathrm{col}(A)$. We assume for case 9 that $\mathrm{col}(A)=0 \pmod{2}$, which ensures 
\begin{equation}
\label{vanish1}
		|( y_1  \textbf{1} )| \cliff 0 .
\end{equation}
Next, we rearrange the last term
\[
 2 |(A^T \vec{x}) \wedge ( y_1  \textbf{1} )|  = 2  y_1 | (A^T \vec{x})\wedge (   \textbf{1} )| ,
\]
and use $\vec{v} \wedge \textbf{1} = \vec{v}$ for all $\vec{v}$, so that 
\[
 2 |(A^T\vec{x}) \wedge ( y_1  \textbf{1} )|  = 2  y_1 |A^T \vec{x}| .
\]
The matrix $A$ is assumed to satisfy $| A^T \vec{x} | = F_U (\vec{x}) +  2 \tilde{F}_U (\vec{x}) \cliff F_U (\vec{x}) $, and using this we have
\[
 2 | (A^T \vec{x} ) \wedge ( y_1  \textbf{1} ) |  = 2  y_1  F_U (\vec{x}) + 4 y_1 \tilde{F}_U(\vec{x}) .
\]
Since $2 y_1 \cliff 0$ and $2 \tilde{F}_U(\vec{x})\cliff 0$, we have by property (4) that $4 y_1 \tilde{F}_U(\vec{x}) \cliff 0$, and so
\begin{equation}
	 2 | (A^T \vec{x} ) \wedge ( y_1  \textbf{1} ) |  \cliff 2  y_1  F_U (\vec{x}).
\end{equation}
Combining the above equations gives
\begin{equation}
\label{BeforeCubic}
	| (A^T\vec{x}) \oplus ( y_1  \textbf{1} ) | \cliff 	F_U (\vec{x}) + 2  y_1  F_U(\vec{x}) .
\end{equation}
The above expressions hold for all unitaries and will be reused later.  We now consider the special case where $F_U$ is homogeneous cubic, and so by property (2) we have $2  y_1  F_U(\vec{x}) \cliff 0$.    This entails 
\begin{equation}
\label{AfterCubic}
	|K_9^T\vec{x} \oplus S^T_9 \vec{y} | =| (A^T\vec{x}) \oplus ( y_1  \textbf{1} ) | \cliff 	| A^T \vec{x} |  ,
\end{equation}
which is the desired result.

Next, we tackle the more general case where $U$ is not a CCZ circuit, but the weighted polynomial $F_U$ still has no linear terms.  Let us consider case 5, and so assume $\mathrm{col}(A)=0 \pmod{2}$ and $\mathrm{col}(B)=0 \pmod{2}$, and set
\[
G_5 =  \left( \begin{array}{c}
K_5   \\ \hline
S_5  \end{array} \right)= \left(  \begin{array}{ccc}
A & B & B  \\  \hline
1 & 1 & 0   \\
0 & 1 & 1  
\end{array}\right) ,
\]
The weight now has three main contributions
\begin{align}
	|K_5^T\vec{x} \oplus S^T_5 \vec{y} | = & | (A^T \vec{x}) \oplus (y_1 \vec{1}) |\\ \nonumber
	& + | (B^T \vec{x}) \oplus ((y_1 \oplus y_2) \vec{1}) | \\ \nonumber
	& + | (B^T \vec{x}) \oplus ( y_2 \vec{1}) | .
\end{align}
We can reuse Eq.~(\ref{BeforeCubic}), and make similar derivations for the $B$ terms, so that
\begin{align}
	|K_5^T\vec{x} \oplus S^T_5 \vec{y} | \cliff & 	F_U(\vec{x}) + 2  y_1  F_U(\vec{x})
 \\ \nonumber
& + F_{V}(\vec{x})+ 2  (y_1 \oplus y_2)  F_{V}(\vec{x})   \\ \nonumber
& + F_{V}(\vec{x}) + 2   y_2  F_{V}(\vec{x})  
\end{align}
The function $F_{V}$ appears twice, but $2 F_{V} \cliff 0$.  Using $y_1 \oplus y_2 = y_1 + y_2 - 2y_1 y_2 $, we deduce that 
\begin{align} \nonumber
	2  ((y_1 \oplus y_2)+y_2)  F_{V} (\vec{x}) & =    (2y_1+4y_2-4y_1y_2)F_{V}(\vec{x})  \\ \nonumber
 & \cliff (2y_1-4y_1y_2)F_{V}(\vec{x}),
\end{align} 
where we have used property (4) in moving to the second line. Combining these observations and regrouping terms gives 
\begin{align} \nonumber
| K_5^T\vec{x} \oplus S^T_5 \vec{y} |   \cliff &  F_U(\vec{x}) + 2 y_1 (F_U(\vec{x})+F_{V}(\vec{x})) \\ \nonumber
   & - 4y_1 y_2 F_V(\vec{x}) .
\end{align}
We know that $F_{V}$ only differs from $F_U$ by cubic terms, and so $F_U(\vec{x})+F_{V}(\vec{x}) = 2L_U(\vec{x})+ 4 Q_U(\vec{x})+4C_U(\vec{x})+4C_V(\vec{x})$ for some linear, quadratic and cubic polynomials.  Since  $2L_U(\vec{x})+ 4 Q_U(\vec{x}) \cliff 0$, we know by property (4) that $(2 y_1)(2L_U(\vec{x})+ 4 Q_U(\vec{x}))\cliff 0$. Although $4C_U(\vec{x})+4C_{V}(\vec{x})$ is not Clifford, it is homogeneous cubic and so by property (2) vanishes when multiplied by $2y_1$.  Therefore, $2 y_1 (F_U(\vec{x})+F_{V}(\vec{x}))\cliff 0$ and so
\begin{equation}
\label{G5termsPRE}
 | K_5^T\vec{x} \oplus S^T_5 \vec{y}  |  \cliff   F_U (\vec{x}) - 4y_1 y_2 F_{V}(\vec{x}) 
\end{equation}
Applying property (3) we have
\begin{equation}
\label{G5terms}
| K_5^T\vec{x} \oplus S^T_5 \vec{y} |  \cliff   F_U(\vec{x}) - 4y_1 y_2 L_U(\vec{x})
\end{equation}
The above has not yet assumed any special properties of the unitary and will be reused later.  Now we use that $F$ has no linear terms, so $L_U(\vec{x})=0$.  This completes the proof of quasitransversality  for case 5.

\begin{table*}
\caption{The matrix $G$ used to design distillation protocols for 11 different cases.  The matrix $G$ is built using $A$ and $B$ matrices and the column vector $c$ that has $c_j = l_j$ where $l_j$ are the linear coefficients of the function $F$.  The appropriate case depends on the properties of $A$ and $B$.  The entries 0 and 1 always denote constant rectangular submatrices of appropriate size to ensure the overall matrix is well formed. When these submatrices are row vectors, we sometimes use the notation $\vec{0}^T$ and $\vec{1}^T$, though not in this table.  The main text provides explicit proofs for cases 9, 5, and then 1.}
\label{TABallCASES}
	\begin{tabular}{| c | c | c |} 
\hline
 & $\mathrm{col}(B) \geq 0$ & $\mathrm{col}(B) \geq 0$\\ 
 & $F_U(\vec{x})=L(\vec{x})+2Q(\vec{x})+4C(\vec{x})$  & $F_U(\vec{x})=2Q(\vec{x})+4C(\vec{x})$   \\ 	 \hline	
 & CASE 1 & CASE 5 \\
$\mathrm{col}(A)$ is even, $\mathrm{col}(B)$ is even
  &  $ \left(  \begin{array}{ccccccccccc}
A & B & B & \vec{c} & \vec{c} & \vec{c} & \vec{c} & 0 & 0 & 0 & 0 \\  \hline
1 & 1 & 0       & 1 & 0 & 0 & 1 & 1 & 0 & 0 & 1  \\
0 & 1 & 1       & 0 & 1 & 0 & 1 & 0 & 1 & 0 & 1 \\
0 & 0 & 0       & 1 & 1 & 1 & 1 & 1 & 1 & 1 & 1 \\
 \end{array}\right) $  &  $ \left(  \begin{array}{ccc}
A & B & B  \\ \hline
1 & 1 & 0  \\
0 & 1 & 1  \\
 \end{array}\right) $ 
 \\
& $\Delta=8$ & $\Delta=0$ \\ \hline
  & CASE 2 & CASE 6 \\
$\mathrm{col}(A)$ is even, $\mathrm{col}(B)$ is odd &
 $ \left(  \begin{array}{cccccccccccc}
A & B & B & \vec{c} & \vec{c} & \vec{c} & \vec{c} & 0 & 0 & 0 & 0 & 0 \\  \hline
1 & 1 & 0 & 1 & 0 & 0 & 1 & 1 & 0 & 0 & 1 & 1  \\
0 & 1 & 1 & 0 & 1 & 0 & 1 & 0 & 1 & 0 & 1 & 1  \\
0 & 0 & 0 & 1 & 1 & 1 & 1 & 1 & 1 & 1 & 1 & 0  \\
 \end{array}\right) $
 &  $ \left(  \begin{array}{ccccc}
A & B & B & 0 & 0 \\ \hline
1 & 1 & 0 & 1 & 0 \\
0 & 1 & 1 & 1 & 1 \\
 \end{array}\right) $
\\ 
 & $\Delta=9$ & $\Delta=1$ \\ \hline
  & CASE 3 & CASE 7 \\
$\mathrm{col}(A)$ is odd, $\mathrm{col}(B)$ is even &
$ \left(  \begin{array}{cccccccccccc}
A & B & B & \vec{c} & \vec{c} & \vec{c} & \vec{c} & 0 & 0 & 0 & 0 & 0 \\  \hline
1 & 1 & 0 & 1 & 0 & 0 & 1 & 1 & 0 & 0 & 1 & 1  \\
1 & 0 & 1 & 0 & 1 & 0 & 1 & 0 & 1 & 0 & 1 & 1  \\
0 & 0 & 0 & 1 & 1 & 1 & 1 & 1 & 1 & 1 & 1 & 0  \\
 \end{array}\right) $
 &  $ \left(  \begin{array}{ccccc}
A & B & B & 0  \\ \hline
1 & 1 & 0 & 1  \\
1 & 0 & 1 & 1  \\
 \end{array}\right) $
 \\ 
& $\Delta=9$ & $\Delta=1$ \\ \hline
  & CASE 4 & CASE 8 \\
$\mathrm{col}(A)$ is odd, $\mathrm{col}(B)$ is odd &
$ \left(  \begin{array}{cccccccccccccc}

A & B & B & \vec{c} & \vec{c} & \vec{c} & \vec{c} & 0 & 0 & 0 & 0 & 0 & 0 & 0 \\  \hline
1 & 1 & 0 & 1 & 0 & 0 & 1 & 1 & 0 & 0 & 1 & 1 & 1 & 0 \\
1 & 0 & 1 & 0 & 1 & 0 & 1 & 0 & 1 & 0 & 1 & 1 & 0 & 1 \\
0 & 0 & 0 & 1 & 1 & 1 & 1 & 1 & 1 & 1 & 1 & 0 & 0 & 0 \\
 \end{array}\right) $
 &  $ \left(  \begin{array}{cccccc}
A & B & B & 0 & 0 & 0\\ \hline
1 & 1 & 0 & 1 & 1 & 0 \\
1 & 0 & 1 & 1 & 0 & 1 \\
 \end{array}\right) $
 \\
& $\Delta=11$ & $\Delta=3$ \\ \hline
	\end{tabular}

\begin{tabular}{| c |c |c |} \hline
	 & $B=0$ & $B=0$ \\
	 & $F_U(\vec{x})=4C(\vec{x})$ & $F_U(\vec{x})=4C(\vec{x})$ \\
	 & $(1,1,\ldots 1) \notin \mathrm{span}[A]$ & $(1,1,\ldots 1) \in \mathrm{span}[A]$ \\ \hline
	 & CASE 9 & CASE 10  \\
$\mathrm{col}(A)$ is even & $ \left( \begin{array}{c} A \\ \hline 1  \end{array} \right) $ & $ \left( \begin{array}{ccc} A & 0 & 0\\ \hline 1 & 1 & 1 \end{array} \right) $ \\ 
& $\Delta=0$ & $\Delta=2$ \\ \hline
	 & CASE 11 &   \\
$\mathrm{col}(A)$ is odd & $ \left( \begin{array}{cc} A & 0 \\ \hline 1 & 1  \end{array} \right) $ &  empty case \\ 
& $\Delta=1$ &   \\ \hline\end{tabular}
\end{table*}

Next, we further broaden the class of unitaries allowing the weighted polynomial to have linear, quadratic and cubic terms.  Though again we take $\mathrm{col}(A)$ and $\mathrm{col}(B)$ to be even.  This is case 1, and the corresponding distillation matrix is 
\begin{equation}
	G_1  =  \left( \begin{array}{c}
K_1   \\ \hline
S_1  \end{array} \right)
=  \left( \begin{array}{cc}
K_5 & \tilde{K}_1   \\ \hline
\rule[-.3\baselineskip]{0pt}{11pt}  S_5 & \tilde{S}_1   \end{array} \right)
\end{equation}
where
\begin{equation}
	\left( \begin{array}{c}
\tilde{K}_1   \\ \hline
\rule[-.3\baselineskip]{0pt}{11pt}  \tilde{S}_1  \end{array} \right) =	\left(  \begin{array}{ccccccccccc}
 \vec{c} & \vec{c} & \vec{c} & \vec{c} & 0 & 0 & 0 & 0 \\  \hline
 1 & 0 & 0 & 1 & 1 & 0 & 0 & 1  \\
 0 & 1 & 0 & 1 & 0 & 1 & 0 & 1 \\
 1 & 1 & 1 & 1 & 1 & 1 & 1 & 1 \\
\end{array}\right) ,
\end{equation}
where $c$ is a column vector with $c_j = l_j$ where $l_j$ is the linear coefficient in the weighted polynomial $F_U$.  So case 1 is similar to case 5, but with an extra 8 columns appended.  We again consider the weight
\begin{equation}
		| K_1^T\vec{x} \oplus S^T_1 \vec{y}  | = 	| K_5^T\vec{x} \oplus S^T_5 \vec{y}  |+ | \tilde{K}_1^T \vec{x} \oplus \tilde{S}_1^T \vec{y} |.
\end{equation}
We can use Eq.~(\ref{G5terms}) to deduce that 
\begin{equation}
		| K_1^T\vec{x} \oplus S^T_1 \vec{y}  | \cliff  F_U (\vec{x}) - 4y_1 y_2 L_U (\vec{x}) + | \tilde{K}_1^T \vec{x} \oplus \tilde{S}_1^T \vec{y}| .
\end{equation}
Since $L(\vec{x})$ is non-zero, we employ the third term to eliminate it.  Evaluating the additional columns using the same methodology we get~(see App.~\ref{Pmat} for details)
\begin{equation}
\label{PmatEq}
	| \tilde{K}_1^T \vec{x} \oplus \tilde{S}_1^T \vec{y}| = 4 \sum_j c_jx_j  y_1 y_2.
\end{equation}
We use that $\vec{c}$ is defined so that $c_j=l_j$ where $l_j$ are the coefficients of $L_U(\vec{x})$, so
\begin{equation}
	| \tilde{K}_1^T \vec{x} \oplus \tilde{S}_1^T \vec{y} | = 4 L_U(\vec{x})  y_1 y_2.
\end{equation}
Therefore, it cancels the linear terms and we have as required
\begin{equation}
		| K_1^T\vec{x} \oplus S^T_1 \vec{y} | \cliff  F(\vec{x}) .
\end{equation}
In the proofs above we used that $| \vec{1} |$ equals either the number of columns in $A$ or $B$, and assumed both these numbers are even.  The remaining cases differ in the number of columns of $A$ and $B$, and this can be accommodated with slight adjustments to additional padding columns.  

It is straightforward to confirm that if $A$ is full rank, then so too is $G$ for every case in Table~\ref{TABallCASES}.  Only in case 10 were some additional columns required to ensure $G$ is full rank.  Much rests on $A$ being full rank.  We explain in Sec.~\ref{PP}  how to cope with rank deficient $A$.  Finally, if the $S$ submatrix of $G$  has nontrivial support on every column, then the corresponding quantum code will have distance 2 or greater.  To see this, note that any $\vec{e}$ with $|\vec{e}|=1$ can only satisfy $S \vec{e}=(0,0,\ldots,0)$ if $S$ has an empty (all zero) column.  This can be seen to hold for all cases 1 to 11 by inspection of Table~\ref{TABallCASES}.

\section{Gate synthesis in the $A$ matrix picture}
\label{secCC}

\subsection{Phase polynomials}
\label{PP}

Numerous papers in the gate-synthesis literature~\cite{amy13,amy14,amy16} make use of phase polynomials, which are an alternative form for weighted polynomials.  We begin by reviewing the formalism of these earlier works, before showing how it fits into the matrix formalism used in defining quantum codes.\begin{defin}
\label{phase_func}
Let $\textbf{a} \in \mathbb{Z}^{2^k}$.  We define a function $P_\mathbf{a}$, which we call a phase polynomial, which can be decomposed as 
\begin{equation}
P_\textbf{a}(\vec{x}) = \sum_{\vec{u} \in \mathbb{Z}_2^{2^k} } a_{\vec{u}} \langle \vec{x}, \vec{u} \rangle \pmod{8} ,
\end{equation}
where we index the elements, $a_\vec{u}$, of $\textbf{a}$ with the label $\vec{u} \in \mathbb{Z}_2^{2^k}$, and make use of the inner product
\begin{equation}
    \langle \vec{x}, \vec{u} \rangle = \bigoplus x_j u_j \pmod{2}. 
\end{equation}
\end{defin}
Notice that the inner product is evaluated $\pmod{2}$, whereas overall the function is defined $\pmod{8}$. The length of the vector $\vec{a}$ is $2^k$, so very large, but its entries are typically sparse. For every weighted polynomial function $F$, there exists~\cite{amy13} a $P_\textbf{a}$ such that $P_\textbf{a}(\vec{x})=F(\vec{x})$ for all $\vec{x}$, which we denote as $P_\textbf{a}=F$.  Once we have a phase polynomial $P_\vec{a}=F$, it is known that $U_F$ can be implemented with $| \vec{a} \pmod{2}|$ uses of $T$, using an established method.  Note that in the expression for the $T$-count we take $\pmod{2}$ before taking the weight.  However, the existence of such phase polynomial representations are not unique.  Amy and Mosca~\cite{amy16} observed that different phase polynomials, with different corresponding $T$-counts, are actually equal functions, taking the same value for all $\vec{x}$.  Specifically, they proved that ancilla-free optimisation of $T$-counts over the $\{$CNOT, $T\}$ basis is equivalent to finding the minimal $| \vec{a} \pmod{2} |$ such that $P_{\vec{a}}=F$.   Denote $\mathcal{V}$ as the set of $\vec{a}$ such that $P_{\vec{a}}(\vec{x})=0$ for all $\vec{x}$.  Since phase polynomials compose linearly $P_{\vec{a}}+P_{\vec{a}'}=P_{\vec{a}+\vec{a}'}$, it follows that if $\vec{a}' \in \mathcal{V}$ then $P_{\vec{a}}=P_{\vec{a}+\vec{a}'}$.  Given an initial $\vec{a}$ such that $P_{\vec{a}}=F$, the optimisation problem is 
\begin{equation}
	\tau[U_{F=P_{\vec{a}}}] := \mathrm{min} 	\{  | (\vec{a}+\vec{a}') \pmod{2} | , \forall \vec{a}' \in \mathcal{V} \}.
\end{equation}
The set $\mathcal{V}$ has a lot of structure.  If $\vec{a}$ and $\vec{a}'$ are in $\mathcal{V}$, then $P_{\vec{a}'+\vec{a}''}=P_{\vec{a}'}+P_{\vec{a}''}=0$ and so $\vec{a}'+\vec{a}''$ is also in $\mathcal{V}$.  Therefore, $\mathcal{V}$ is an Abelian group using addition in $\mathbb{Z}_8$.  Since, the weight is evaluated modulo 2, we are actually interested in $\mathcal{V}_2 = \{ \vec{a} \pmod{2} : \vec{a} \in \mathcal{V} \}$ which also forms a group, though this time over $\mathbb{Z}_2$ and so $\mathcal{V}_2$ is a vector space.  Amy and Mosca showed that $\mathcal{V}_2$ corresponds to the codewords of the punctured Reed-Muller code over $2^n-1$ bits and with order $(n-4)$, which is more succinctly denoted by $\mathcal{RM}(n,n-4)^*$.  Therefore, the $T$-count optimisation is equivalent to minimum weight decoding over Reed-Muller codes.  Unfortunately, no efficient optimal decoders are known.  For small circuits,  brute force optimisation is feasible.  For larger circuits, we may have to settle for suboptimal methods.   We return to the optimality question in the following subsections.

Next, we explain how phase polynomials relate to quantum codes with quasitransversal gates.  Here we show the following
\begin{lem}
\label{constructG}
Let $U$ be a unitary with weighted polynomial $F$, and let $P_\vec{a}$ be a phase polynomial satisfying $F=P_\vec{a}$.  It follows that unitary $U$ has a gate-synthesis matrix $A$ with $\mathrm{col}(A)= | \vec{a} \pmod{2} |$.  Specifically, $A$ is a matrix where the column vector $\vec{u}$ appears once if and only if $a_{\vec{u}} =1 \pmod{2}$.
\end{lem}

This construction of $A$ will be central to our entire framework.  Before we give the general proof let us consider an example.

\begin{exmp}
\label{CS_gate_synthesis}
The control-S unitary $U_{CS}$ has weighted polynomial $F=2x_1 x_2$.  This is equal to $P_{\vec{a}}=   x_1 + x_2 - (x_1 \oplus x_2) =   x_1 +  x_2 +  7(x_1 \oplus x_2) \pmod{8}$ .  In other words, $\vec{a}$ is a vector where $a_{(1,0)}=1$,  $a_{(0,1)}=1$, $a_{(1,1)}=7$ and all other elements are zero.  Therefore, the vectors $(1,0)$, $(0,1)$, and $(1,1)$ satisfy $a_{\vec{u}}=1 \pmod{2}$, and we construct $A_{CS}$ using these three columns vectors
\begin{equation}
\label{ACS}
    A = \left( \begin{array}{ccc}
     1 & 0 & 1 \\
     0 & 1 & 1 \\
 \end{array} \right).
\end{equation}
One can verify that
\begin{equation}
	| A^T\vec{x} |=  (x_1 \oplus x_2) + x_1 + x_1 .
\end{equation}
Therefore, $F=P_{\vec{a}}=| A^T\vec{x} |+2 \tilde{F}$ where $\tilde{F}=3( x_1 \oplus  x_2 )=3x_1+3x_2+2x_1x_2$, and so  $| A^T\vec{x} | \cliff P_{\vec{a}}$.
\end{exmp}

We begin our proof of the lemma by observing that when calculating weights of vectors, the order of elements is irrelevant and we can consider $A$ to be a set of column vectors $\{ \vec{u} \} \in A$.  The weight $|A^T \vec{x}|$ can then be re-expressed as 
\begin{align}
    |A^T \vec{x}| & = \sum_{\vec{u} \in A} \langle \vec{u} , \vec{x} \rangle  \\ \nonumber
    & = \sum_{\vec{u} \in \mathbb{Z}^{2^k}_2 } v_{\vec{u}} \langle \vec{u} , \vec{x} \rangle \\ \nonumber 
    & = P_{\vec{v}}
\end{align}
where in the second line we have extended the sum over the whole domain by introducing the indicator vector $\vec{v} \in \mathbb{Z}_2^{2^k}$ with elements $v_{\vec{u}}=a_{\vec{u}} \pmod{2}$.  Therefore, there exists some binary vector $\vec{\tilde{a}}$ such that $\vec{v} = \vec{a} + 2 \vec{\tilde{a}}$, and so
\begin{equation}
	P_{\vec{v}}=	P_{\vec{a}}+2P_{\vec{\tilde{a}}}.
\end{equation}
The additional $2P_{\vec{\tilde{a}}}$ corresponds to some weighted polynomial $2\tilde{F}$, and so $P_{\vec{v}} \cliff 	P_{\vec{a}}$.  Combined with  $|A^T \vec{x}|=P_{\vec{v}}$, we deduce $|A^T \vec{x}|\cliff 	P_{\vec{a}}$, which completes the proof of Lem.~\ref{constructG}.  

Let us recap how this relates to the gate synthesis problem.  If $A$ is a gate-synthesis matrix for $U$, then assuming $A$ is full rank, it defines a trivial quantum code, with $K=A$ and $S$ being empty. Therefore Thm.~\ref{thm_main} shows that we can use Clifford operations and $\mathrm{col}(A)$ $T$-gates to prepare the magic state $U \ket{+}^{\otimes k} $, and $U$ can be injected into an algorithm using teleportation~\cite{Zhou00}. Although here there is no error suppression, since $\epsilon_{\mathrm{out}}=O(\epsilon)$ for a trivial code.  This offers a different perspective on gate-synthesis.

Extension to the case where $A$ is rank deficient follows from a Clifford equivalence argument. Consider a rank deficient $A$ that is a gate-synthesis matrix for unitary $U$.  We show $U$ is Clifford equivalent to a unitary $U' \otimes \id$ where $U'$ acts on a smaller number of qubits and has a full rank gate-synthesis matrix $A'$. We show this by considering how a matrix $A$ constructed from function $P_\vec{a}$, acts under Clifford equivalences. Consider Cliffords composed of CNOT gates that act as $E_J \ket{\vec{x}} = \ket{J^T \vec{x}}$ where $J$ is an invertible square matrix over $\mathbb{Z}_2$. It follows that
\begin{align}
	E_J^\dagger	U_{P_\vec{a}} E_J \ket{\vec{x}}& =E_J^ \dagger	U_{P_\vec{a}}  \ket{J^T \vec{x}} \\ \nonumber
& =E_J^\dagger	\omega^{P_\vec{a}(J^T \vec{x})}  \ket{J^T \vec{x}} \\ \nonumber
& =	\omega^{P_\vec{a}(J^T \vec{x})}  \ket{\vec{x}}
\end{align}
and so we have a new phase polynomial
\begin{equation}
	P_\vec{b}( \vec{x})	= P_\vec{a}( J^T \vec{x}).
\end{equation}
Using the definition of phase polynomials we have
\begin{equation}
	\sum_{\vec{u}} b_{\vec{u}} \langle \vec{u} , \vec{x} \rangle	=	\sum_{\vec{u}} a_{\vec{u}} \langle \vec{u} , J^T\vec{x} \rangle.
\end{equation}
Since the inner product satisfies $\langle \vec{u} , J^T\vec{x} \rangle = \langle J\vec{u} , \vec{x} \rangle$, we have
\begin{equation}
	\sum_{\vec{u}} b_{\vec{u}} \langle \vec{u} , \vec{x} \rangle	=	\sum_{\vec{u}} a_{\vec{u}} \langle J \vec{u} , \vec{x} \rangle.
\end{equation}
 Since $J$ is invertible we may change variables $\vec{u} \rightarrow J \vec{u}$ on the left hand side so that
\begin{equation}
	\sum_{\vec{u}} b_{J\vec{u}} \langle J \vec{u} , \vec{x} \rangle	=	\sum_{\vec{u}} a_{\vec{u}} \langle J \vec{u} , \vec{x} \rangle,
\end{equation}
so we clearly see $a_{\vec{u}} = b_{J \vec{u}} $.  Therefore, the new gate synthesis matrix contains the column $J\vec{u}$ whenever $A$ contains $\vec{u}$.  That is, we have mapped
\begin{equation}
	A \rightarrow JA .	
\end{equation}
We can always find an $J$ such that 
\begin{equation}
    J A = \left(\begin{array}{c} 
A' \\
0 \\
\end{array}
  \right)  ,
\end{equation}
where $A'$ is full rank and the corresponding unitary acts on a smaller number of qubits than for $A$.   As such, herein we will always consider full rank $A$ matrices. 

\subsection{Lempel factorisation}
\label{Bmatrixfast}

For our protocols to offer substantial improvement we need to know that there are many cases where the majority of the $T$-count is due to CCZ gates.   For this reason, we introduced a decomposition $U=VW$ where $W$ must be composed solely of CCZ gates, but $V$ is otherwise an arbitrary gate in $\mathcal{D}_3$.  The remainder $V$ has $T$-count $\tau[V]$, and minimising over all such decompositions gives the quantity $\mu[U]$ as defined in Def.~\ref{muDef}.  Earlier, we saw that circuits with small ratio $\mu[U] / \tau[U]$ offer the best resource saving for synthillation.  However, we have not yet seen how to determine $\mu[U]$.  Thm.~\ref{circuits} asserted that for $k$ qubit unitaries $\mu[U] \leq k+1$ and the exact value of $\mu[U]$ can be efficiently found.  We prove this theorem here. Before proceeding, we define an equivalence relation.
\begin{defin}
\label{PauliEq}
	Given a weighted polynomial $F$,  if there exists a homogeneous cubic function $4C$ and $2\tilde{F} \cliff 0$ such that $F=F' + 2\tilde{F}+ 4C$ we write $F \pauli F'$.
\end{defin}
In other words, if $F \pauli F'$ then
\begin{align}
	U_F & = V_{F' + 2\tilde{F}} W_{4C} \\ \nonumber
	& = V_{F'}V_{2\tilde{F}} W_{4C} , 
\end{align}
where $W_{4C} \in \mathcal{D}_3^C$.  From this we may infer $\mu[U_F ]=\mu[V_{F'}]$.  The very definition of $\mu$ can be recast in these terms as $\mu[U_F]:=\mathrm{min} \{ \tau[V_{F'}] | F' \pauli F \}$.  Lastly, $\cliff$ is finer than $\pauli$, which means that if $F \cliff F'$ then $F \pauli F'$.  This can be verified by setting $4C=0$. 

We work in a matrix picture and find a $B$, which is the optimal gate-synthesis matrix for $V$ and has $\mathrm{col}(B) = \tau[V] = \mu[U]$.  Another useful matrix representation is the following.
\begin{defin}
\label{QmatDefin}
Let $F_{U}$ be the weighted polynomial for unitary $U$ with coefficients $l_i, q_{i,j}, c_{i,j,k}$ as in Eq.~(\ref{function_form}).  We define the quadratic-matrix for $U$ as follows: $Q$ is a binary symmetric matrix such that $Q_{i,j}=Q_{j,i}=q_{i,j} \pmod{2}$ for $i \neq j$ and $Q_{i,i}=l_i \pmod{2}$.
\end{defin}
With these definitions, finding $\mu[U]$ can be recast as a matrix factorisation problem
\begin{lem}
\label{LempelReduction}
Let unitary $U$ have quadratic-matrix $Q$ and unitary $V$ have gate synthesis matrix $B$.  It follows that $Q=B \cdot B^T \pmod{2}$ if and only if $F_U \pauli F_{V}$.  Therefore, $\mu[U] = \mathrm{min} \{ \mathrm{col}(B) | Q=B \cdot B^T \pmod{2} \}$.
\end{lem}
Before proving the lemma, we discuss it consequences.  This matrix factorisation problem turns out to be a well known problem, which can be efficiently solved using Lempel's factorization algorithm~\cite{lempel75}.  This minimal construction of $B$ has $\mathrm{rank}[Q]+1$ columns if the diagonal entries of $Q$ are all zero, and otherwise the minimal $B$ has $\mathrm{rank}[Q]$ columns. The rank of $Q$ can not exceed the number of columns in $Q$, and the number of columns  equals the number of qubits that the unitary acts on.  Therefore, for a $k$ qubit unitary, there exists a  suitable $B$ with at most $k+1$ columns, and this entails $\tau[V] \leq k+1$ and so $\mu[U] \leq k+1$.   Therefore,  Lem.~\ref{LempelReduction} and Lempel's factorization algorithm directly entails Thm.~\ref{circuits}.  The Supplementary material provides a mathematica script for Lempel's algorithm and a faster variant we found~\cite{CampSupp16}.

We originally posed $\mu$ as a double optimisation problem, where one of the optimisations, evaluating $\tau$, appears to be hard.  Nevertheless finding $\mu$ is easy, and it is informative to delve deeper into the comparison of these problem.  We reviewed earlier that finding the optimal $T$-count was equivalent to decoding the Reed-Muller code $\mathcal{RM}(n,n-4)^*$.  In contrast, Lempel and Seroussi~\cite{seroussi80} showed matrix factorisation is equivalent to decoding the Reed-Muller code $\mathcal{RM}(n,n-3)^*$.  Consequently, the Reed-Muller code $\mathcal{RM}(n,n-3)^*$ can be efficiently decoded, whereas $\mathcal{RM}(n,n-4)^*$ cannot.

Now we commence the proof.  
\begin{proof}
By construction, there exist integer $\tilde{q}_{i,j}$ and $\tilde{l}_{j}$ such that  $q_{i,j}=Q_{j,i}+2 \tilde{q}_{i,j} $ and $l_{j}=Q_{j,j}+2 \tilde{l}_{j} $.  We apply these substitutions to the expansion of $F_U$ (recall Eq.~\eqref{function_form}) to obtain
\begin{align}
	F_U (\vec{x}) = &  \sum_{i} Q_{i,i} x_i + 2\sum_{i<j} Q_{i,j} x_i x_j \\ \nonumber
& + 2 ( \sum_i \tilde{l}_{i} x_i  + \sum_{i < j } 2\tilde{q}_{i,j} x_i x_j ) \\ \nonumber
& + 4 \sum_{i<j<k} c_{i,j,k} x_i x_j x_k .
\end{align}
The second line is Clifford, and so
\begin{align}
	F_U (\vec{x}) \cliff & \sum_{i} Q_{i,i} x_i + 2\sum_{i<j} Q_{i,j} x_i x_j   \\ \nonumber
& + 4 \sum_{i<j<k} c_{i,j,k} x_i x_j x_k.
\end{align}
Next, we rearrange the first line as follows
\begin{align}
	F_U (\vec{x}) \cliff &  \left(\sum_{i,j} Q_{i,j} x_i x_j \right) \\ \nonumber
& +  4 \sum_{i<j<k} c_{i,j,k}  x_i x_j x_k.
\end{align}
Observe that the linear term is slightly hidden, but still present since for binary variables we have $x_i x_i = x_i$.  Also, the quadratic terms still carry a prefactor two because we now sum over all $i$ and $j$ and so double count every $i \neq j$ contribution.  Being ambivalent over cubic terms we write
\begin{equation}
	F_U (\vec{x}) \pauli \sum_{i,j} Q_{i,j} x_i x_j .
\end{equation}
Now we turn our attention to $B$. Evaluating the weight of $B^T \vec{x}$,
\begin{equation}
	 | B^T \vec{x} | = \sum_h [ \oplus_i B_{i,h} x_i \pmod{2} ].	
\end{equation}
The switch from modular to standard arithmetic gives 
\begin{align}
\label{BigBeq}
 | B^T \vec{x} |   \pauli &\sum_h \Bigg[ \left( \sum_{i} B_{i,h} x_i \right) \\  \nonumber
 & - 2 \left( \sum_{i < j} B_{i,h} B_{j,h} x_i x_j \right) \Bigg] ,
\end{align}
where cubic terms and higher degree terms have been dropped due to the $\pauli$ relation.   Adding terms of the form $4x_i x_j$ gives a Clifford equivalent function, and so we can replace the $-2$ with $+2$. Notice that the second set of terms is over $i<j$. Extending the sum over all $i \neq j$ simply double counts every entry so
\begin{equation}
	2 \sum_h \sum_{i<j} B_{i,h} B_{j,h} x_i x_j =	  \sum_h  \sum_{i \neq j} B_{i,h} B_{j,h} x_i x_j.
\end{equation}
Looking at the first set of terms,  using that $B$ and $\vec{x}$ are binary, we have
\begin{align}
	\sum_{h,i} B_{i,h} x_i & = \sum_{h,i} B_{i,h} B_{i,h} x_i x_i \\
& = \sum_{h,j=i} B_{i,h} B_{j,h} x_i x_j .
\end{align}
In the second line we introduce the dummy variable $j$, but the sum is fixed $j=i$.  This dummy variable serves to clarify the connection between this equation and the quadratic terms above.  This allows us to simplify Eq.~(\ref{BigBeq}) by merging the linear and quadratic contributions into a single sum
\begin{equation}
		| B^T \vec{x} | \pauli \sum_h \sum_{i,j} B_{i,h} B_{j,h} x_i x_j  .
\end{equation}
The sum over $h$ is simply matrix multiplication so that 
\begin{equation}
		| B^T \vec{x} | \pauli \sum_{i,j} [ B \cdot B^T ]_{i,j} x_i x_j  .
\end{equation}
From the transitivity of equivalence relations we deduce $| B^T \vec{x} | \pauli F_U (\vec{x})$ if and only if 
\begin{equation}
	\sum_{i,j} Q_{i,j} x_i x_j  \pauli \sum_{i,j} [ B \cdot B^T ]_{i,j} x_i x_j.
\end{equation}
Clearly, this is satisfied if $Q=B \cdot B^T$.  Furthermore, neither side carries any cubic terms, and since both matrices are binary the coefficients are either 0 or 1 for linear terms and 0 or 2 for quadratic terms, and so the relation only holds if $Q=B \cdot B^T$.  This is last step in reducing our problem to matrix factorisation.
\end{proof}

\subsection{A fast algorithm for finding $T$-counts}

Here we establish some tools for effectively finding good gate-synthesis matrices.  Amy and Mosca~\cite{amy16} showed that for any $k$-qubit unitary in $\mathcal{D}_3$, the cost of optimal gate-synthesis scales asymptotically as $O(k^2)$.  Specifically, that for large $k$, we have $\tau[U] \leq \frac{1}{2}k^2 -1$.  However, for large circuits, finding optimal solutions with known algorithms is slow.  We need to resort to suboptimal solutions, and it is unclear how far these will deviate from the worst case scaling.  A previously proposed approach is the $T_{PAR}$ algorithm~\cite{amy14}, though it comes with no promise on the maximum $T$ count.  Writing out an explicit circuit for $U$ there are at most $O(k^3)$ gates, which is dominated by the $n$ choose 3 possible CCZ gates, which leads a naive decomposition using $O(k^3)$ $T$-gates. Neither existing algorithm is connected to the  $O(k^2)$ scaling of optimal solutions.

This section gives a proof of Thm.~\ref{subOPT}, which shows that there exists a polynomial time algorithm that outputs a gate sequence using no more $T$-gates than $\sim \frac{1}{2}k^2$.  While there is no promise that our algorithm gives an optimal output, our solution is fast and obeys the same scaling as optimal solutions.  

Our proof rests on the following lemma
\begin{lem}
\label{decompLemma}
Let $U_k \in \mathcal{D}_3$ act on $k$ qubits.  We can in polynomial time find a decomposition of $U_k$ into $U_k=\tilde{U}_{k}  U_{k-1}$ such that $\tilde{U}_{k}, U_{k-1} \in \mathcal{D}_3$ with $U_{k-1}$ acting nontrivially on $k-1$ qubits.  Furthermore, in polynomial time we can find a circuit that realises $\tilde{U}_k$ using no more than $(k+1)$ $T$-gates.
\end{lem}
The above decomposition entails 
\begin{align}
		\tau[U_k]	& \leq   \tau[\tilde{U}_{k}] + \tau[ U_{k-1}] \\ \nonumber
& \leq   (k+1)+ \tau[ U_{k-1}] 
\end{align}
We proceed iteratively, invoking the above procedure down to a $c$ qubit problem.  We choose  $c$ to be constant and sufficiently small that optimal decoding is tractable.  The $T$ gates used in this decomposition are bounded so that
\begin{align}
	\tau_{\mathrm{fast}} & \leq \tau[U_{c}]  +\sum_{j=c+1}^k (j+1) \\ \nonumber
& \leq \tau[U_{c}]  +\frac{ k^2 + 3k- (c^2+3c)}{2}
 \end{align}
For $c=4$, the decoding problem is simple as there are only two phase polynomials to check.  Furthermore, for 4 qubits it is known~\cite{amy16} that $\tau$ never exceeds 7, so
\begin{align}
	\tau_{\mathrm{fast}}  & \leq  7 + \frac{ k^2 + 3k -28}{2}. \\
& = \frac{k^2 + 3 k - 14 }{2}.
 \end{align}
Therefore, Thm.~\ref{subOPT} follows from Lem.~\ref{decompLemma}.  We remark that $c=4$ was chosen for simplicity, but one should use the largest value of $c$ for which an $\mathcal{RM}(c,c-4)^*$ decoder runs in acceptable time.

\begin{proof}
Let us now prove the above lemma and use $F_k$ for the weighted polynomial corresponding to unitary $U_k$.   We can always sort the terms of the weighted polynomial so that
\begin{align}
\label{eq_func_split}
	F_k(\vec{x}) =&  f_{k-1}(\vec{x}') + 2 x_k  g(\vec{x}') + l_k x_k    ,
\end{align}
where $\vec{x}'$ equals $\vec{x}$ with the last element removed, so that
\begin{equation}
	\vec{x} = \left( \begin{array}{c}
 	\vec{x}' \\ \nonumber
	x_k
 \end{array}
  \right).
\end{equation}
The function $f_{k-1}$ collects all terms that are independent of $x_k$.  The term $2 x_k  g$ collects all terms involving $x_k$ and at least one other variable. The last term $l_k x_k$, with $l_k \in \mathbb{Z}_8$, captures whether $x_k$ appears alone.  Since $g$ is multiplied by $2 x_k$, cubic terms in $g$ will vanish modulo 8. In other words, $g$ is only defined upto a CCZ circuit.  Therefore, Thm.~\ref{thm_prots} ensures we can efficiently find $B$ with $\mathrm{col}(B) \leq k $ such that $|B^T\vec{x}'| \pauli g(\vec{x}') $, and so  $2x_k |B^T \vec{x}' | \cliff 2x_k g(\vec{x}') $.  Note that the relevant inequality is $\mathrm{col}(B) \leq k $ rather than $\mathrm{col}(B) \leq k+1 $ because $g$ is defined over $k-1$ variables rather than $k$. Next, we recall the modular identity $2| \vec{u} \wedge \vec{v} |=  |\vec{u}| + |\vec{v}| - |\vec{u} \oplus \vec{v}|$, and set $\vec{u}=x_k\vec{1}$ and $\vec{v}=B^T \vec{x}'$ to infer
\begin{align}
	2x_k  |B^T \vec{x}' | &= 2 | (x_k \vec{1}) \wedge (B^T \vec{x}') | \\ \nonumber
& = x_k | \vec{1} | +  | B^T \vec{x}' | -  | (x_k \vec{1}) \oplus (B^T \vec{x}') | .
\end{align}
Substituting this into Eq.~(\ref{eq_func_split}) and by virtue of $2x_k g(\vec{x}') \cliff 2x_k |B^T \vec{x}' |$, we have
\begin{align}
\label{eq_func_split2}
	F_k(\vec{x}) \cliff &  \tilde{F}_k(\vec{x}') + F_{k-1}(\vec{x}) ,
\end{align}
where we have defined new functions $\tilde{F}_k$ and $F_{k-1}$ that collect terms as follows
\begin{align}
	F_{k-1}(\vec{x})  & = f_k(\vec{x}')+  | B^T \vec{x}' | , \\ 
	\tilde{F}_k(\vec{x}') & =  (l_k +  | \vec{1} | )x_k  -  | (x_k \vec{1}) \oplus (B^T \vec{x}') |    \label{Fk_1} .
\end{align}
We now define the decomposition $U_k = \tilde{U}_k U_{k-1}$ so that $ \tilde{U}_k$ is associated with function $\tilde{F}_k$ and $ U_{k-1}$ is associated with function $F_{k-1}$.  Since $F_{k-1}$ is independent of $x_k$, the unitary $ U_{k-1}$ acts nontrivially on no more than $k-1$ qubits.  It remains to find a decomposition of $ \tilde{U}_{k}$ in terms of $T$ gates, which we do by finding a gate-synthesis matrix.  We define $l \in \{ 0,1\}$ so that $l = l_k +  | \vec{1} | \pmod{2}$, and construct the matrix
\begin{equation}
	A=\left( \begin{array}{c c}
B & 0 \\
\vec{1}^T & l 	
 \end{array}
 \right)	.
\end{equation}
This satisfies $|A^T\vec{x}|=|(B^T \vec{x}') \oplus (x_k \vec{1})|+l x_k $ and  combined with Eq.~(\ref{Fk_1}) entails $F_{k-1}(\vec{x}) \cliff |A^T \vec{x}|$.  Furthermore, $\mathrm{col}(A)=\mathrm{col}(B)+1$.  If $l=0$ then the last column is unnecessary, but to find the upper bound we consider the worst case where $l=1$. Above we saw $\mathrm{col}(B) \leq k$, which entails $\mathrm{col}(A)\leq k+1$.  Since $A$ is the gate-synthesis matrix for $\tilde{U}_k$, we deduce $\tau[ \tilde{U}_k ] \leq k+1$.  \end{proof}

This proves Lem.~\ref{decompLemma}, which in turn proves Thm.~\ref{subOPT}

\subsection{Optimal synthesis of controlled-unitaries}
\label{optimal_cont_U}
 
Here we consider controlled-unitaries.  For this subclass we find it is especially easy to find gate-synthesis matrices using some ideas from the previous section.  
\begin{theorem}[The controlled-unitary theorem]
\label{contU_them}
Let $U \in \mathcal{D}_3$ be a $k$-qubit unitary of the controlled-unitary with target unitary $U_t^2$
\begin{equation}
	U = \id \otimes \kb{0}{0} + U_t^2 \otimes \kb{1}{1},
\end{equation}
where $U_t \in  \mathcal{D}_3$.  It follows that
\begin{equation}
	\tau[U] = \begin{cases}
 2 \mu[U_t] & \textrm{ if } 	 \mu[U_t]  \textrm{ is even;} 	\\
 2 \mu[U_t]  +1 & \textrm{ if } 	 \mu[U_t]  \textrm{ is odd.} 	
 \end{cases}
\end{equation}
Furthermore, we can efficiently find an optimal gate-synthesis matrix for $U$.
\end{theorem}
Notice the important role again played by $\mu$, which emerges because $U_t$ is squared.  From the definition of $\mu$, we have a decomposition $U_t = V_t W_t$ where $\tau[V_t]=\mu[U_t]$ and $W_t$ is composed of CCZ gates.  Since CCZ gates square to the identity it follows that $U_t^2 = V_t^2$. Therefore, we proceed by showing how to implement a controlled-$V_t^2$ unitary, which equals $U$.  In our functional language, $U$ has weighted polynomial $F(\vec{x})=2x_k g(\vec{x}')$ where $g(\vec{x})$ is the weighted polynomial for $V_t$.

The previous section established that using Lempel's factorisation method for finding $V_t$ and a gate synthesis matrix $B$ for $V_t$ with $\mathrm{col}(B) = \tau[V_t]$ and $|B^T\vec{x}'| \cliff g(\vec{x}')$.  Using $l = | \vec{1} | \pmod{2}$ where $| \vec{1} | = \mathrm{col}(B) $, we construct 
\begin{equation}
\label{Aopt}
	A=\left( \begin{array}{c c c}
B & B & 0 \\
\vec{1}^T & \vec{0}^T & l 	
 \end{array}
 \right)	.
\end{equation}
We evaluate
\begin{align}
	|A^T \vec{x}| & = |(B^T \vec{x}')\oplus(\vec{1})|+ |B^T \vec{x}'|+ l x_k.
\end{align}
and using the modular identity and simplifying, we find
\begin{align}
	|A^T \vec{x}| & \cliff  2 x_k |B^T \vec{x}'| .
\end{align}
Using that $|B^T\vec{x}'| \cliff g(\vec{x}')$, we deduce $|A^T \vec{x}| \cliff  2 x_k g(\vec{x}') = F(\vec{x})$.  This is entails that $A$ is a gate synthesis matrix for $U$.  Clearly, $\mathrm{col}(A) \leq 2\mathrm{col}(B)+1$ and using $\mathrm{col}(B)\leq k$ we arrive at $\mathrm{col}(A) \leq 2k+1  $.  As promised, these controlled-unitaries require at most $(2k+1)$ $T$-gates.  We claimed that this is an optimal solution.   Showing this is a tedious variant of the above, so we relegate it to App.~\ref{converse}.  

A very simple example is a CS gate, for which $B=(1)$ and so we have the $A_{CS}$ matrix already given in Eq.~(\ref{ACS}). We give a new example  here.  
\begin{exmp}
\label{TofHash}
Let $\mathrm{tof}_{\#}$ be a pair of Toffolis with a single control in common.
The phrase $\mathrm{tof}_{\#}$ was coined in Ref.~\cite{gossett98} where the gate appears naturally in Shor's algorithm.  The gate $\mathrm{tof}_{\#}$ is Clifford equivalent to a pair of CCZ gates with associated weighted polynomial
\begin{align}
	F_{\#}(\vec{x}) & = 4x_1 x_2 x_5 + 4x_3 x_4 x_5 \\ \nonumber
& = 2x_5 ( 2 x_1 x_2 + 2 x_3 x_4) \\ \nonumber
& = 2 x_5 g( x_1 , x_2, x_3, x_4)
\end{align}
where in last line we defined $g( x_, x_2, x_3, x_4) :=  2 x_1 x_2 + 2 x_3 x_4$.   We must find a $B$ so that $|B^T\vec{x}| \pauli  2 x_1 x_2 + 2 x_3 x_4$, which is equivalent to solving the factorisation problem $Q= B \cdot B^T$ where 
\begin{equation}
	Q = \left( \begin{array}{cccc}
0 & 1 & 0 & 0 \\
1 & 0 & 0 & 0 \\
0 & 0 & 0 & 1 \\
0 & 0 & 1 & 0 
\end{array} \right).	
\end{equation}
Applying Lempel's factorisation method (see Sec.~\ref{Bmatrixfast}) gives
\begin{equation}
 B =	\left( \begin{array}{ccccc}
0 & 0 & 0 & 1 & 1 \\
0 & 0 & 1 & 1 & 0 \\
0 & 1 & 1 & 1 & 1 \\
1 & 0 & 1 & 1 & 1 
\end{array} \right).
\end{equation}
Since $\mathrm{col}(B)=5=1\pmod{2}$, we have $l=1$ and so define
\begin{align}
\label{Amat_TofHahs}
	A & = \left( \begin{array}{c c c}
B & B & 0 \\
  \vec{1}^T & \vec{0}^T	 & 1 	
 \end{array}
 \right)	\\ 
& = \left( \begin{array}{ccccccccccc}
0 & 0 & 0 & 1 & 1 & 0 & 0 & 0 & 1 & 1 & 0 \\
0 & 0 & 1 & 1 & 0 & 0 & 0 & 1 & 1 & 0 & 0 \\
0 & 1 & 1 & 1 & 1 & 0 & 1 & 1 & 1 & 1 & 0 \\
1 & 0 & 1 & 1 & 1 & 1 & 0 & 1 & 1 & 1 & 0 \\
1 & 1 & 1 & 1 & 1 & 0 & 0 & 0 & 0 & 0 & 1 
 \end{array}
 \right).
\end{align}
As promised, $|A^T \vec{x}| \cliff F_{\#}(\vec{x}) $, and we have $\mathrm{col}(A)=11$ and so $\tau[\mathrm{tof}_{\#}] = 11$. \end{exmp}
We will later reuse the $\mathrm{tof}_{\#}$ example as a case study for synthillation. 

\subsection{Subadditivity of $T$-count}
\label{subAdd}

Here we share a curious observation on the behavior of the optimal $T$-count.  Given a tensor product of two unitaries, $U = U_1 \otimes U_2$, one has directly that $\tau[U] \leq  \tau[U_1]+\tau[U_2]$ simply by treating the two problems as separate.  Since the circuits act on distinct blocks of qubits, there are no obvious places that $T$ gates cancel in the decomposition and so one might expect additivity to hold $\tau[U] = \tau[U_1]+\tau[U_2]$.  Here we give examples and general classes of strictly subadditive behaviour where $\tau[U] < \tau[U_1]+\tau[U_2]$.  Practically, this entails resource savings by preparing joint batches of unitaries.

The most general form of our observation is the following
\begin{theorem}[Subadditivity theorem]
\label{SubaddThm}
	Let $U_{1}, U_2 \in \mathcal{D}_3$ where $U_1$ is a circuit composed of CCZ gates and $\tau[U_{1}], \tau[U_2] > 0$. If $\tau[U_1]=1\pmod{2}$ then $\tau[U_1 \otimes U_2] \leq \tau [ U_1 ] + \tau [U_2]-1 < \tau[U_1] + \tau [U_2] $.
\end{theorem}
We prove this by directly constructing a gate synthesis matrix for $U_1 \otimes U_2$.  Let $A_1$ and $A_2$ be optimal gate-synthesis matrices for $U_1$ and $U_2$.  Since $\tau[U_2]>0$, the matrix $A_2$ has at least one column. We use $\vec{z}$ to denote the first column of $A_2$ so that $A_2=[\vec{z} , A_{*}]$ where $A_{*}$ denotes the remaining columns.  Using this column vector we define $R$ as $R= \vec{z} (\vec{1}^T) = (\vec{z}, \vec{z}, \ldots \vec{z})$ where $\vec{1}$ is the all unit vector such that $\mathrm{col}(R)=|\vec{1}|=\mathrm{col}(A_1)=\tau[U_1]$.  We now construct the matrix
\begin{equation}
	A = \left( \begin{array}{cc}
	A_1 & 0 \\
	R   & A_{*} 
 \end{array}
\right)	.
\end{equation}
The idea is that the submatrix $R$ acts as a substitute for the first column of $A_2$, and so this column can be trimmed off leaving $A_{*}$. There is nothing unique about the first column.  We pick it out merely for concreteness.  Notice that $\mathrm{col}(A)=\mathrm{col}(A_1)+\mathrm{col}(A_{*})$, and so $\mathrm{col}(A)=\mathrm{col}(A_1)+\mathrm{col}(A_2)-1=\tau[U_1]+\tau[U_2]-1$.  It remains to be shown that $A$ is a gate-synthesis matrix for $U_1 \otimes U_2$.  We again calculate $|A^T \vec{x}|$ and partition $\vec{x}$ into $\vec{x}'$ and $\vec{x}''$, so that $|A^T \vec{x}| = |(A_1^T \vec{x}') \oplus (R^T \vec{x}'')| + |A_{*}^T\vec{x}''|$.  The standard switching  of arithmetic gives 
\begin{align}
	 |A^T \vec{x}| =& |A_1^T \vec{x}'| + | R^T\vec{x}'' | \\ \nonumber
& - 2| (A_1^T \vec{x}') \wedge (R^T \vec{x}'') | + |A_{*}^T\vec{x}''|.
\end{align}
We have 
\begin{align}
		(R^T \vec{x}'') &= (\vec{z} (\vec{1}^T))^T \vec{x}''  \\ \nonumber
	&=(\vec{1} (\vec{z}^T))\vec{x}''  \\ \nonumber
	&=\vec{1}  (\vec{z}^T \vec{x}'')  
\end{align}
and so $|R^T \vec{x}''|= |\vec{1}| (\vec{z}^T \vec{x}'')  = \tau[U_1](\vec{z}^T \vec{x}'') $.  The assumption $\tau[U_1]=1\pmod{2}$ entails $|\vec{1}|=1\pmod{2}$ and so $|R^T \vec{x}''| \cliff \vec{z}^T \vec{x}'' $. At this point, we have
\begin{align}
	 |A^T \vec{x}| \cliff & |A_1^T \vec{x}'| + \vec{z}^T\vec{x}'' \\ \nonumber
& - 2| (A_1^T \vec{x}') \wedge (R^T \vec{x}'') | + |A_{*}^T\vec{x}''|.
\end{align}
Next, we observe that $\vec{z}^T\vec{x}'' +  |A_*^T \vec{x}''| = |A_2^T \vec{x}''|$ so that
\begin{align}
	 |A^T \vec{x}| &\cliff |A_1^T \vec{x}'| + | A^T_2\vec{x}'' |  - 2| (A_1^T \vec{x}') \wedge (R^T \vec{x}'') | .
\end{align}
Next, we show the last term vanishes.  We can evaluate this wedge product by considering two cases.  If $\vec{z}^T \vec{x}''= 0 $ then $(A_1^T \vec{x}') \wedge (R^T \vec{x}'')=\vec{0}$.  Whereas if $\vec{z}^T \vec{x}''= 1 $ then $(A_1^T \vec{x}') \wedge (\vec{1})=A_1^T \vec{x}'$.  These two cases are succinctly captured by
\begin{equation}
	2| (A_1^T \vec{x}') \wedge (R^T \vec{x}'') | =(2\vec{z}^T \vec{x}'')  |A_1^T \vec{x}'|.
\end{equation}
We now use that $|A_1^T \vec{x}'|$ is equivalent to some homogenous cubic polynomial because $U_1$ is a CCZ circuit.  As we have seen before, homogenous cubic polynomials carry a prefactor 4 and so vanish $\pmod{8}$ when multiplied by $(2\vec{z}^T \vec{x}'')$.  Therefore,
\begin{equation}
	 |A^T \vec{x}| \cliff |A_1^T \vec{x}'| + |A^T_2\vec{x}'' | ,
\end{equation}
which shows $A$ is a gate synthesis matrix achieving the same effect as $A_1$ and $A_2$ and so $U_1$ and $U_2$.  This completes the proof.  

The simplest example is
\begin{exmp}
	Let $U_{\mathrm{1CCZ}+T} = U_{\mathrm{1CCZ}} \otimes T$ where $U_{\mathrm{1CCZ}}$ is a single CCZ gate.  It is well known that $\tau[U_{\mathrm{1CCZ}}]=7$ and this can be achieved with gate-synthesis matrix
\begin{equation}
\label{gate_synthesis_matrix_1CCZ}
		A_{\mathrm{1CCZ}} = \left( \begin{array}{cccccccc}
1 & 1 & 0 & 1 & 1 & 0 & 0   \\
1 & 0 & 1 & 1 & 0 & 1 & 0  \\
0 & 1 & 1 & 1 & 0 & 0 & 1    \\	
 \end{array} \right).
\end{equation}
For $T$ the gate-synthesis matrix is simply $A_T = (1)$ and so $z=(1)$ and $A_{*}$ is an empty matrix.  Following our construction, we have that a gate-synthesis matrix for $U_{\mathrm{1CCZ}+T}$ is
\begin{equation}
		A_{\mathrm{1CCZ}+T} = \left( \begin{array}{cccccccc}
1 & 1 & 0 & 1 & 1 & 0 & 0   \\
1 & 0 & 1 & 1 & 0 & 1 & 0  \\
0 & 1 & 1 & 1 & 0 & 0 & 1    \\	
1 & 1 & 1 & 1 & 1 & 1 & 1  \\	
 \end{array} \right).
\end{equation}
Therefore, $\tau[U_{\mathrm{1CCZ}+T}]=7 < \tau[U_{\mathrm{1CCZ}}]+\tau[T] = 8$.
\end{exmp}
We also have more complex examples
\begin{exmp}
	Let $U_{\mathrm{2CCZ}} = U_{\mathrm{1CCZ}} \otimes U_{\mathrm{1CCZ}}$ where $U_{\mathrm{1CCZ}}$ is a single CCZ gate.  The first column of $A_{\mathrm{1CCZ}}$ above is $\vec{z}=(1,1,0)^T$, and so
\begin{equation}
\label{gate_synth_2CCZ}
		A_{\mathrm{2CCZ}} = \left( \begin{array}{ccccccccccccccc}
1 & 1 & 0 & 1 & 1 & 0 & 0 & 0 & 0 & 0 & 0 & 0 & 0  \\
1 & 0 & 1 & 1 & 0 & 1 & 0 & 0 & 0 & 0 & 0 & 0 & 0 \\
0 & 1 & 1 & 1 & 0 & 0 & 1 & 0 & 0 & 0 & 0 & 0 & 0   \\	
1 & 1 & 1 & 1 & 1 & 1 & 1 & 1 & 0 & 1 & 1 & 0 & 0   \\
1 & 1 & 1 & 1 & 1 & 1 & 1 & 0 & 1 & 1 & 0 & 1 & 0  \\
0 & 0 & 0 & 0 & 0 & 0 & 0 & 1 & 1 & 1 & 0 & 0 & 1    \\	
 \end{array} \right).
\end{equation}
Therefore, $\tau[U_{\mathrm{2CCZ}}] \leq 13 < 2 \tau[U_{\mathrm{1CCZ}}]= 14$.
\end{exmp}
This use of subadditivity can be extended by noting that
if both $U_1$ and $U_2$ are CCZ circuits with odd $T$-count, then $U_1 \otimes U_2$ is also a CCZ circuit with odd $T$-count.  This enables the proof to be iterated so that we have
\begin{corollary}
	Let $\{ U_1, U_2, \ldots U_N \}$ be a set of circuits each composed from CCZ gates with $\tau[U_j] = 1 \pmod{2}$ for all $j$.   It follows that for $U = \otimes U_j$ we have $\tau[U]\leq \left(  \sum_j \tau[U_j] \right)-(n-1)$.
\end{corollary}
For instance, given $n$ copies of $U_{\mathrm{1CCZ}}$ we have $\tau[ U_{\mathrm{1CCZ}}^{\otimes N} ] = 7N+(N-1)=6N+1$.  We see the cost per gate asymptotically approaches 6 rather than 7.  Similarly, the $\mathrm{tof}_{\#}$ gate of Example.~\ref{TofHash} is a CCZ circuit with $\tau[\mathrm{tof}_{\#}]=11$ and so $\tau[\mathrm{tof}_{\#}^{\otimes N}] \leq 10 N+1$.

\section{Applications}
\label{sec_Applications}

This section draws upon the set of techniques developed to present  specific synthillation protocols.  The case studies are chosen to most clearly demonstrate the general techniques.

\subsection{Toffoli gates}

The simplest application of synthillation is for implementing a Toffoli gate, or equivalently a CCZ gate denoted $U_{\mathrm{1CCZ}}$.  It is well known that a CCZ gate can be realised using 7 $T$-gates, and by considering the possible phase polynomial representations we deduce this is the lowest $T$-count possible without ancilla-assistance, and so $\tau[U_{\mathrm{1CCZ}}]=7$.  We remind the reader that $\tau$ was defined as the ancilla-free $T$-count, and throughout have used the phrase gate-synthesis synonymously with this ancilla-free notion of gate-synthesis.  With the aid of ancilla, a Toffoli can be realised using only 4 $T$-gates~\cite{jones13b}, and we return to this point in the discussion section.

 If we use synthillation to prepare a single Toffoli gate, we have the following protocol
\begin{exmp}
\label{1Toff}
Synthillation for a single CCZ gate $U_{\mathrm{1CCZ}}$.  We have $\tau[U_{\mathrm{1CCZ}}]=7$ and clearly $\mu[U_{\mathrm{1CCZ}}]=0$.  The problem falls into case 11 of Table~\ref{TABallCASES} and we use the gate-synthesis matrix from  Eq.~(\ref{gate_synthesis_matrix_1CCZ}), so that
\begin{equation}
		G = \left( \begin{array}{cccccccc}
1 & 1 & 0 & 1 & 1 & 0 & 0 & 0   \\
1 & 0 & 1 & 1 & 0 & 1 & 0 & 0 \\
0 & 1 & 1 & 1 & 0 & 0 & 1 & 0   \\ \hline
1 & 1 & 1 & 1 & 1 & 1 & 1 & 1	
 \end{array} \right).
\end{equation}
Therefore, it uses 8 $T$-states of error rate $\epsilon$ to perform a CCZ gate with probability and error rates:
\begin{align} \nonumber
			p_{\mathrm{suc}} & = 1-8 \epsilon+56 \epsilon^2-224 \epsilon^3+560 \epsilon^4-896 \epsilon^5 + O(\epsilon^6) \\ \nonumber
  \epsilon_{\mathrm{out}}    & = 28 \epsilon^2-168 \epsilon^3+476 \epsilon^4-784 \epsilon^5+784 \epsilon^6 + O(\epsilon^7)
\end{align}
Full expressions available in Supplementary Material~\cite{CampSupp16}.
\end{exmp}

The above protocol performs identically to that of Eastin~\cite{eastin13} and Jones~\cite{jones13b}, which were shown to outperform all previous protocols.  The research undertaken here began as an attempt to recast these Toffoli protocols in the $G$ matrix formalism, and then extended this insight to the whole family of $\mathcal{D}_3$ gates.  However, our techniques can improve over these single Toffoli protocols.  By producing a batch of single Toffoli states,  we can exploit the subadditivity shown in Sec.~\ref{subAdd}.  We learned that $\tau[U_{\mathrm{1CCZ}}^{\otimes N}] \leq 6 N+1$.  This $T$-count is odd, and so the synthillation protocol again falls into case 11 of Table~\ref{TABallCASES}. Therefore, it uses $n=6N+2$ noisy $T$-states to output $N$ error suppressed Toffoli gates.   Asymptotically, this approaches 6 per Toffoli, and so gives approximately a $25\%$ reduction in resources over the Eastin~\cite{eastin13} and Jones~\cite{jones13b} protocols.  In general, the success probability is determined by the span of $S$ via Eq.~(\ref{MacWillpsuc}) and so
\begin{align}
	p_{\mathrm{suc}} &=\frac{1}{2}\left( 1+(1-2\epsilon)^{6N+2} \right)	
\end{align}
The error rate $\epsilon_{\mathrm{out}}$ can be exactly calculated for any particular $N$, but does not have such a simple form. However, an upper bound on  $\epsilon_{\mathrm{out}}$ is readily available.  We know the $\vec{e}=(0,0,\ldots 0)$ vector corresponds to no errors.  Therefore, we obtain an upper bound on the output error by summing over all nontrivial even weight bit strings and renormalizing
\begin{align}
	\epsilon_{\mathrm{out}}  &\leq 1 - \frac{2(1-\epsilon)^{6N+2}}{1+(1-2\epsilon)^{6N+2}}
\end{align}
We go into more detail for the $N=2$ protocol.

\begin{exmp}
\label{2Toff}
Synthillation for two Toffoli gates $U_{\mathrm{2CCZ}}=U_{\mathrm{1CCZ}}^{\otimes 2}$.  It is known $\tau[U_{\mathrm{2CCZ}}]=13$ and clearly $\mu[U_{\mathrm{2CCZ}}]=0$.  The problem falls into case 11 of Table~\ref{TABallCASES} and we may use gate-synthesis  matrix $A_{2CCZ}$ as given in Eq.~(\ref{gate_synth_2CCZ}), so that
\begin{equation}
	G_{2CCZ}=\left( \begin{array}{cccccccccccccccc}
1 & 1 & 0 & 1 & 1 & 0 & 0 & 0 & 0 & 0 & 0 & 0 & 0 & 0   \\
1 & 0 & 1 & 1 & 0 & 1 & 0 & 0 & 0 & 0 & 0 & 0 & 0 & 0   \\
0 & 1 & 1 & 1 & 0 & 0 & 1 & 0 & 0 & 0 & 0 & 0 & 0 & 0   \\	
1 & 1 & 1 & 1 & 1 & 1 & 1 & 1 & 0 & 1 & 1 & 0 & 0 & 0   \\
1 & 1 & 1 & 1 & 1 & 1 & 1 & 0 & 1 & 1 & 0 & 1 & 0 & 0   \\
0 & 0 & 0 & 0 & 0 & 0 & 0 & 1 & 1 & 1 & 0 & 0 & 1 & 0   \\	 \hline
1 & 1 & 1 & 1 & 1 & 1 & 1 & 1 & 1 & 1 & 1 & 1 & 1 & 1
 \end{array} \right).
\end{equation}
Therefore, it uses 14 noisy $T$-states of error rate $\epsilon$ to perform two CCZ gates with probability and error rates:
\begin{align} \nonumber
		p_{\mathrm{suc}}= &   1-14 \epsilon+182 \epsilon^2-1456 \epsilon^3+8008 \epsilon^4 + O(\epsilon^5) \\ \nonumber
		\epsilon_{\mathrm{out}}  & = 91 \epsilon^2+182 \epsilon^3-7021 \epsilon^4-28812 \epsilon^5 + O(\epsilon^6)
\end{align}
See Supplementary Material~\cite{CampSupp16} for further details.
\end{exmp}
Above we focused on comparison with Eastin~\cite{eastin13} and Jones~\cite{jones13b}, but it is also important to reflect on the advantage over traditional distill-then-synthesize methods.  Using BHMSD and gate-synthesis one obtains $N$ error-suppressed Toffoli gates using $(3N+8) \tau [U_{\mathrm{NCCZ}}] =(3N+8)(6N+1)$ resources, which is about a factor 3 worse than synthillation.  In this comparison, we have even allowed distill-then-synthesize to leverage subadditivity.

\begin{figure*}
	\includegraphics{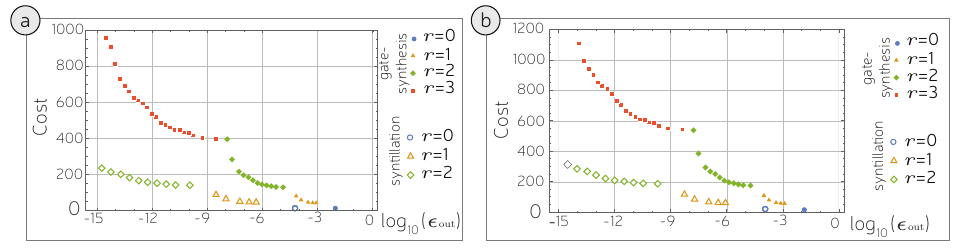}
	\caption{Resource cost measured by expected number of raw  ($\epsilon =0.001$) magic states consumed in implementing: (a) $\tau[U_{2\#}]$ and (b) $\tau[U_{3\#}]$. Costs are plotted against $\epsilon_{\mathrm{out}}$, the error rate on the implemented gate. We compare using synthillation and distill-then-synthesize (BHMSD with optimal gate-synthesis). Both protocols use $r$ precursor rounds of BHMSD, where $r$ is chosen to ensure a target error rate is reached. We implement single gate and so do not take advantage of the batch discount due to subadditivity.}
\label{numerics}  
\end{figure*}

\subsection{Control-$S$ gates}

Here we consider the problem of implementing many control-$S$ gates, which we call CS for short.  Specifically, we set $U_{NCS}=U_{CS}^{\otimes N}$.  We choose this task primarily for pedagogical purposes as it provides clear exposition of several of our techniques and relates to the counterintuitive circuit decomposition shown in Fig~(\ref{fig:circuits}d).  For a single control-$S$ gate it is well known that $\tau[U_{CS}]=3$ with gate-synthesis matrix $A_{CS}$ introduced in Eq.~(\ref{CS_gate_synthesis}). The subadditivity theorem (Thm.~\ref{SubaddThm}) does not apply here, and for $N=2$ we have solved the optimal decoding problem to verify that $\tau[U_{CS}^{\otimes 2}]=6$.  We therefore conjecture that these gate behave additively, so that $\tau[U_{CS}^{\otimes N}]=3N$, and proceed on this assumption. Next, we evaluate $\mu[ U_{CS}^{\otimes N}]$ using the method presented in Sec.~\ref{Bmatrixfast}.  We note that $U_{NCS}$ corresponds to a weighted polynomial
\begin{equation}
	F_{NCS}(\vec{x}) = 2 \sum_{j} x_{2j-1}x_{2j}.
\end{equation}
	The coefficients of this function define a $2N$-by-$2N$ symmetric matrix (recall Def.~\ref{QmatDefin}) 
\begin{align}
	Q & = \left( \begin{array}{ccccccc}
 	0 & 1 & 0 & 0 &    & 0 & 0  \\ 
 	1 & 0 & 0 & 0 &  \ldots  & 0 & 0  \\   
 	0 & 0 & 0 & 1 &   \ldots & 0 & 0  \\   
 	0 & 0 & 1 & 0 &    & 0 & 0  \\   
& \vdots & \vdots &  &   \ddots &  &   \\ 
 	0 & 0 & 0 & 0 &    & 0 & 1  \\ 
 	0 & 0 & 0 & 0 &    & 1 & 0  \\ 
 \end{array}
\right) ,
\end{align}
which can be compactly written as
\begin{equation}
	Q = X \otimes \id_N	,
\end{equation}
where $X$ is the Pauli-$X$ operator, $\id_N$ is the $N$-by-$N$ identity matrix, and $\otimes$ is the tensor product.  Clearly, $Q$ is full rank, so $\mathrm{rank}[Q]=2N$ and has zero entries on the diagonal.  Therefore, Lempel factorization yields a $B$ satisfying $Q = B. B^T \pmod{2}$ with $\mathrm{col}(B)=2N+1$.  Therefore, $\mu[U_{NCS}]=2N+1$.  We observe $F_{NCS}$ has no linear terms, and consult Table.~\ref{TABallCASES} to construct synthillation protocols using $n$ resources, where
\begin{equation}
		n = \begin{cases}
 		7N+3,  \textrm{ using case 6 for all even } N, \\
		7N+5,  \textrm{ using case 8 for all odd } N.
 \end{cases}
\end{equation}
In both cases, the cost approaches 7 per CS gate. But even $N$ is slightly better, so we use  that case for the following analysis.  The success probability of synthillation depends only on the lower submatrix of $G$ and is found (using Eq.~(\ref{MacWillpsuc})) to be
\begin{equation}
	p_{\mathrm{suc}} = \frac{1}{4}\left( 1+ (1-2\epsilon)^{4N+2}+ 2(1-\epsilon)(1-2\epsilon)^{5N+1}\right)
\end{equation}
It is informative to provide an upperbound on the error out by again making the pessimistic assumption that all $\vec{e} \neq (0,0,\ldots 0)$ lead to output errors, and so
\begin{equation}
	\epsilon_{\mathrm{out}} \leq 1 - \frac{4(1-\epsilon)^{7N+3}}{1+ (1-2\epsilon)^{4N+2}+ 2(1-\epsilon)(1-2\epsilon)^{5N+1}}
\end{equation}
Let us again consider a concrete example
\begin{exmp}
\label{2NCS}
Synthillation for two CS gates $U_{\mathrm{2CS}}=U_{\mathrm{CS}}^{\otimes 2}$.  It is known that $\tau[U_{\mathrm{2CS}}]=6$ and $\mu[U_{\mathrm{2CS}}]=5$.     Performing Lempel factorisation we find
\begin{equation}
\label{eq_B2CS}
	B_{2CS}= \left( \begin{array}{ccccc}
 		0 & 0 & 0 & 1 & 1 \\
		0 & 0 & 1 & 1 & 0 \\
        0 & 1 & 1 & 1 & 1 \\ 
		1 & 0 & 1 & 1 & 1 
 \end{array}
\right)	.
\end{equation}
Therefore, there exists a decomposition $U_{\mathrm{2CS}}=V W$ where $\tau[V]=5$ and the weighted polynomial for this circuit is
\begin{align}
	F_V(\vec{x}) & = 2(x_1 x_2 +x_3 x_4) + 4(x_1 x_2 x_4 + x_1 x_2 x_4) \\ \nonumber
 & \cliff | B_{2CS}^T\vec{x} | .
\end{align}
We see $V$ differs from $U$ by the addition of two CCZ gates, so this is the $U=V W$ decomposition shown earlier in Fig.~(\ref{fig:circuits}d).  
  We use $B_{2CS}$, two instances of the gate-synthesis  matrix $A_{CS}$ from Eq.~(\ref{ACS}) and case 6 of Table~\ref{TABallCASES} to construct
\begin{equation}
	G_{2CS}=\left( \begin{array}{cccccc|ccccc|ccccc|cc}
 1 & 0 & 1 & 0 & 0 & 0 &   0 & 0 & 0 & 1 & 1   & 0 & 0 & 0 & 1 & 1 &    0 & 0 \\
 0 & 1 & 1 & 0 & 0 & 0 &   0 & 0 & 1 & 1 & 0   & 0 & 0 & 1 & 1 & 0 &    0 & 0 \\
 0 & 0 & 0 & 1 & 0 & 1 &   0 & 1 & 1 & 1 & 1   & 0 & 1 & 1 & 1 & 1 &    0 & 0 \\ 
 0 & 0 & 0 & 0 & 1 & 1 &   1 & 0 & 1 & 1 & 1   & 1 & 0 & 1 & 1 & 1 &    0 & 0 \\ \hline
 1 & 1 & 1 & 1 & 1 & 1 &   1 & 1 & 1 & 1 & 1   & 0 & 0 & 0 & 0 & 0 &    1 & 0 \\
 0 & 0 & 0 & 0 & 0 & 0 &   1 & 1 & 1 & 1 & 1   & 1 & 1 & 1 & 1 & 1 &    1 & 1 \\
 \end{array} \right). \nonumber
\end{equation}
The vertical lines are merely guides to show the submatrices composing $G_{2CS}$.
Therefore, it uses 18 noisy $T$-states of error rate $\epsilon$ to perform two CS gates with:
\begin{align} \nonumber
		p_{\mathrm{suc}}= &   1-18\epsilon + O(\epsilon^2) \\ \nonumber
		\epsilon_{\mathrm{out}}  & = 45\epsilon^2+294\epsilon^3-603\epsilon^4-20880\epsilon^5+O(\epsilon^6) 
\end{align}
See Supplementary Material~\cite{CampSupp16} for further details.
\end{exmp}

Let us compare to the traditional distill-then-synthesize methods.  To synthesize $N$ CS gates uses $3N$ distilled $T$ states. Therefore, one first uses BHMSD to distil $3k+8 \rightarrow k$, setting $k=3N$ we find the total cost is approximately $9N+8$ noisy $T$ states. The asymptotic cost is 9 per CS gate, and so higher than the 7 per CS gate achieved by synthillation. Furthermore, distill-then-synthesize carries an additive $+8$ cost and so approaches the asymptotic limit considerably slower than synthillation with an additive $+3$ cost.  Our success probability and error out are also comparably better than in the distill-then-synthesize paradigm.  As always, synthillation is beneficial. Although, in this example the resource savings are less than the factor 3 achieved by the best instances of synthillation.  However, our motivation here has been principally educational purposes, and establishing groundwork for the next section.

\subsection{The $U_{N \#} $ family}

Here we consider a family of circuits that extends Toffoli and Tof$_{\#}$ introduced in Example.~\ref{TofHash}.  We define $U_{N \#} $ to be the $2N+1$ qubit unitary composed of $N$ CCZs, which all share exactly one control in common.  Therefore, the CCZ gate is $U_{1\#}$,  the earlier Tof$_{\#}$ gate is $U_{2 \#}$, and then we have newly defined gates $U_{3 \#}$ and onwards.  With common control qubit $k$, the weighted polynomial is
\begin{equation}
	F(\vec{x}) = 4 x_k \sum_{j=1}^{N} x_{2j-1}x_{2j}	.
\end{equation}
Remember from the last section that the many control-$S$ unitary $U_{NCS}$ is described by the weighted polynomial $F_{NCS}(\vec{x})=2\sum_{j=1}^{N} x_{2j-1}x_{2j}$, and so $F_{N \# } =  x_k 2F_{NCS}$.  

We see the $U_{N\#}$ family is closely related to $U_{NCS}$.  Indeed, $U_{N\#}$ can be considered a control-$U_{NCS}^2$ gate.  There is some redundancy here, as $U_{N\#}$ is a control-$U^2$ for any $U = U_{NCS} V $ for any $V$ composed of CCZ gates since $V^2 = \id$. This redundancy was exploited in Sec.~\ref{optimal_cont_U} to find optimal decompositions for general control-unitaries.  Leveraging these results, we have that $\tau[U_{N\#}]=2 \mu[ U_{NCS} ] + 1$.  The last subsection showed  $ \mu[ U_{NCS} ]=2N+1$ and so $\tau[U_{N\#}]=4N + 3 $.  The circuit is composed of $N$ CCZ gates each with $\tau[U_{1CCZ}]=7$ (or $\sim 6$ using subadditivity) and so gains are made over naive gate-synthesis.  Since $\tau[U_{N\#}]$ is odd, this again falls into case 11 of Table~\ref{TABallCASES} and so the synthillation cost is $n=\tau[U_{N\#}]+1=4N + 4$.  Since $\tau[U_{N\#}]$ is odd we may again use subadditivity to obtain a discount if a batch $U_{N\#}^{\otimes m}$ gates is needed. From Thm.~\ref{SubaddThm}, it follows that
\begin{align}
		\tau[U_{N\#}^{\otimes m}] & \leq m(4N + 3)-(m-1) , \\ \nonumber
& = m(4N + 2)+1 .
\end{align}
Distillation of a batch again falls into case 11, and so costs $n= m(4N + 2)+2$ resources per attempt.  As in the previous two case studies, we know
\begin{align}
   p_{\mathrm{suc}} & =\frac{1}{2}\left( 1+(1-2 \epsilon)^{m(4N + 2)+2} \right)	 \\ \nonumber
	\epsilon_{\mathrm{out}}& \leq 1 - \frac{2(1-\epsilon)^{m(4N + 2)+2}}{1+(1-2\epsilon)^{m(4N + 2)+2}}.
\end{align}
We give the simplest example more explicitly.
\begin{exmp}
The unitary $\tau[U_{2\#}]$ has weighed polynomial $F_{2 \#}(\vec{x}) = 4 x_5 (x_1 x_2 + x_3 x_4)=2x_5 F_{2CS}(\vec{x})$. It is the same unitary as considered in example~\ref{TofHash}.  Using the gate-synthesis matrix from Eq.~(\ref{Amat_TofHahs}) with $\tau[U_{2\#}]=11$ and case 11 of Table~\ref{TABallCASES}, we have a synthillation cost of 12 and
\begin{equation}
	G_{2\#}  = \left( \begin{array}{cccccccccccc}
0 & 0 & 0 & 1 & 1 & 0 & 0 & 0 & 1 & 1 & 0 & 0 \\
0 & 0 & 1 & 1 & 0 & 0 & 0 & 1 & 1 & 0 & 0 & 0 \\
0 & 1 & 1 & 1 & 1 & 0 & 1 & 1 & 1 & 1 & 0 & 0 \\
1 & 0 & 1 & 1 & 1 & 1 & 0 & 1 & 1 & 1 & 0 & 0 \\
1 & 1 & 1 & 1 & 1 & 0 & 0 & 0 & 0 & 0 & 1 & 0 \\ \hline
1 & 1 & 1 & 1 & 1 & 1 & 1 & 1 & 1 & 1 & 1 & 1 \\ 
 \end{array}
 \right).	
\end{equation}
Explicit calculation yields
\begin{align}  \nonumber
	p_{\mathrm{suc}} & =  1-12 \epsilon+132 \epsilon^2-880 \epsilon^3+3960 \epsilon^4+O(\epsilon^5) \\ \nonumber
	\epsilon_{\mathrm{out}} & = 66 \epsilon^2+132 \epsilon^3-3678 \epsilon^4-15240 \epsilon^5+O(\epsilon^6)  \nonumber
\end{align}
\end{exmp}
Using the above example we performed numerics finding the expected number of raw magic states needed to distill $\tau[U_{2\#}]$ states of error rate $\epsilon_{\mathrm{target}}$ or less.  We see $\sim 3$ advantage as we expect, though remark that this advantage would be greater if compared against naive rather than optimal gate-synthesis.  We also perform the analysis for $\tau[U_{3\#}]$ and see very similar behaviour but with all costs shifted slightly upwards. In the analysis for all these data points, we use single-shot protocols that do not exploit the subadditivity of preparing batches of gates.  Using subadditivity, all the data points will drop in cost by between $8\%$ and $16\%$.

\section{Discussion}
\label{discuss}

Clifford gates must be supplemented with gates from the third, or higher, level of the Clifford hierarchy in order to achieve universal quantum computation. Here we presented a general framework for preparing purified resource (magic) states that enable multiqubit unitaries from the third level of the hierarchy.  Because this framework combines gate-synthesis and one round of magic state distillation we call it synthillation.  Our first major result is to show large resource savings over the best existing schemes.  For a broad class of circuits, including all circuits composed of control-control-$Z$ gates, the magic state cost of synthillation is approximately the same as gate-synthesis.  Therefore, for these circuits we get a free round of quadratic error suppression, reducing resource costs by roughly a third.  

Optimal solutions of the multiqubit gate synthesis problem are believed to be difficult.  Our second major result is to provide a near-optimal and efficient gate-synthesis algorithm, making use of Lempel's matrix factorisation algorithm. This algorithm efficiently finds $k$-qubit gate decompositions with a cost that scales as $O(k^2)$ in the worst case.  This scaling matches the upper bound of optimal gate-synthesis. Although, for problems that are far from worst case instances, our solution could be far from optimal. We also showed that Lempel factorisation helps with the design of synthillation protocols and can be leveraged to efficiently solve optimal gate-synthesis for the special case of controlled-unitaries.   

Remarkably, we also highlighted that strict subadditivity of $T$-count is possible and in fact commonplace.  Practically, this enables a resource saving on implementing batches of unitaries.  From a fundamental perspective this has a pleasing parallel with other resource theories.  

Having recapped on our results, we address several natural discussion points. We have used BHMSD (Bravyi-Haah magic state distillation) and ancilla-free gate-synthesis as our benchmarks for the distil-then-synthesize paradigm.  But there are other protocols.  First, we discuss ancilla-assisted gate-synthesis.  Recall that $\mathrm{tof}^*$, the gate shown in Fig.~\ref{fig:circuits}c, needs 4 $T$ gates to synthesize without ancilla. It has been shown~\cite{selinger13,jones13b} that ancilla can convert $\mathrm{tof}^*$ into the Toffoli, which needs 7 $T$ gates to synthesize without ancilla.  While this is a remarkable drop in cost, Jones~\cite{jones13b} showed that his Toffoli distillation protocol~\cite{jones13b,eastin13} is more efficient than using BHMSD and then synthesizing $\mathrm{tof}^*$.  The Jones and Eastin protocols are special cases of synthillation, so our approach retains its lead against ancilla-assisted gate-synthesis.  Furthermore, while synthillation can be optimised for general circuits, we know of no general set of tools for ancilla-assisted gate-synthesis.  Understanding the power of ancilla-assistance is an obvious direction for future research.  Another natural question is whether gains can be made by using synthillation to prepare a $\mathrm{tof}^*$ resource, and then using ancilla-assistance to convert it into a Toffoli.  The cost of synthillation does depend on the ancilla-free gate-synthesis cost, but it also depends on other factors.   Because $\mathrm{tof}^*$ is not comprised solely of control-control-Z gates, this increases the synthillation cost, and it turns out it is best to stick with synthillation of the pure Toffoli.  However, there may be other instances were ancilla-assisted techniques pair well with synthillation. 

Another way we can alter the benchmark is to look at distillation routines other than BHMSD.  The most interesting alternative is the multi-level protocol of Jones~\cite{Jones13} as it has superior resource scaling.  Multi-level distillation works best at low error rates.  When targeting error rates around the $10^{-9}$ to $10^{-20}$ range, the level-2 distiller can be used.  Jones gave higher level distillers, but they excel at below $\sim 10^{-20}$ error rates.  Any quantum computer targeting below $10^{-20}$ will be colossal in scale, so let us set that aside as distant future technology.  The level-2 distiller, takes $5k^3 + 24k^2+ 32k$ noisy $T$ states of error rate $\epsilon$ and outputs $k^3$ distilled $T$ states of error rate $O(\epsilon^4)$.  In the large $k$ limit, the cost per output is 5 whereas for two rounds of BHMSD it is 9, so one can expect a factor $\sim 1.8$ improvement over BHMSD.  This is not as large as the factor 3 reduction that can be obtained by using synthillation composed with one round of BHMSD. Furthermore, multilevel distillation must output very large batches (large $k$) to achieve this boost.  This can lead to wasteful oversupply of magic states, even when running quantum computers at maximum clock rates~\cite{fowler12b,Ogorman16}.  Synthillation can be more parsimonious than multi-level distillation, and does not depend on efficiencies of scale to achieve this.  If $O(\epsilon^8)$ or greater error suppression is needed, then synthillation can be composed with multi-level distillation.  A last comment on multi-level distillation is that no full space-time resource analysis, including Clifford costs, has yet been performed for this protocol.  Because multi-level distillation uses bigger jumps in error suppression, it is unclear whether it can fully exploit resource scaling~\cite{RHG01a,fowler13,Ogorman16} (called balanced investment in Ref.~\cite{Ogorman16}), which plays an important role in minimising full resource costs. 

While synthillation was our main focus, we made several contributions relating to optimal gate-synthesis.  It remains to be seen how our general solver compares against the $T_{PAR}$ algorithm~\cite{amy14}.  We also cannot say, without knowing the optimal solution, how close these algorithms come to optimality.  Assuming finding an optimal solution is a hard problem~\cite{seroussi80}, we would like to know how close an efficient algorithm can get to optimality and what the easy instances are.  Clearly, more investigation is needed.  Also of interest is a more comprehensive understanding of subadditivity and whether our results here can be strengthened. 

The exact multiqubit gate synthesis problem considered here concerns the third level of the Clifford hierarchy.  The mathematics lends itself to extensions to higher levels of the hierarchy~\cite{amy16}, and we have found the same holds for synthillation.  We do not report those results here as it appears there are no practical savings to be made.  Synthillation protocols become rapidly more expensive as the hierarchy is ascended, losing all practical merit.  The situation is akin to the work of Landahl and Cesare~\cite{landahl13} where they sought single qubit gates from higher in the hierarchy using codes that fit neatly in the $G$-matrix formalism.  They saw some success for the first few additional levels of the hierarchy, but the costs escalated rapidly.  There has been recent progress on the single-qubit higher-level problem, but using swap gadgets that do not seem to fit neatly within the $G$-matrix framework~\cite{duclos15,campbell16}.  A cohesive understanding of swap gadgets and $G$-matrices appears the best route up the hierarchy.  

The authors are also fond of qudit ($d$-level rather than 2 level) variants of these questions.  We have learnt much about qudit magic state distillation~\cite{anwar12,campbell12,campbell14,watson15,dawkins15} and the qudit Clifford hierarchy~\cite{Howard12}.  However, very little is known yet about qudit gate-synthesis.

Our analysis so far has assessed cost in terms of raw magic states consumed, neglecting resources associated with Cliffords and the underlying error correction code.  Such full resource counts are architecture specific and substantial research projects in their own right~\cite{fowler13,Ogorman16}.  It has recently been argued that a CNOT costs approximately $\sim 1/50$ the value of a $T$-gate obtained via two rounds of distillation~\cite{maslov16}. The dominant Clifford cost in synthillation will be the CNOTs that compose the encoder unitary, and for CNOT circuits there are techniques for minimising resources costs~\cite{patel03,maslov07}.  However, for any particular synthillation problem there exist many equivalent encoder unitaries, each corresponding to a different solution of a matrix completion problem.  An open question here is how to search this equivalence class for the most resource efficient solution.  In a full resource analysis, judicious scaling of code distances~\cite{RHG01a,fowler13,Ogorman16} means that more resources are allocated during the last round of magic state distillation.  Our synthillation protocol focuses on improving final round performance and so targets the known bottleneck point in a full resource analysis.   The full cost of a round of distillation is much more than a factor of 3, and so synthillation may offer a much larger reduction in real terms.

We have taken another step toward minimal resource quantum computing and shown that interesting things can emerge when one delves into the interface of magic states and gate-synthesis.  Individually, both topics have contributed significantly to our understanding of quantum computation, but perhaps they should not be separate topics at all.

\subsection{Acknowledgements}

We acknowledge support by the EPSRC (grant EP/M024261/1). We thank Ben Brown, Joe O'Gorman, Matthew Amy, and Dmitri Maslov for comments on the manuscript.  Thanks to Anqi Gong for spotting some (post-publication) errors in case 6 of Table I and Example IV.3, both of which are fixed in this version.

\appendix

\section{Clifford hierarchy proofs}
\label{App_CliffHier}

Here we show that for all $U \in \mathcal{D}_3$, we have that $U^2$ is Clifford and $U$ is in the third level of the Clifford hierarchy.   First, we observe that if two diagonal unitaries $U_1$ and $U_2$ are members of $\mathcal{C}_j$, then the products $U_1U_2$ also belongs to $\mathcal{C}_j$. Therefore, to confirm a group of unitary operators are all members of $\mathcal{C}_j$, it suffices to check a set of generator are inside $\mathcal{C}_j$.  Given a $U_F \in \mathcal{D}_3$, we have $U_F^2=U_{2F}$ where
\begin{equation}
	2F(\vec{x})	= 2\sum_{i} l_i x_i  +  4 \sum_{i<j}  q_{i,j} x_i x_j \pmod{2},
\end{equation}
and 
\begin{equation}
	U_{2F} = \left( \bigotimes_{i} S_i^{l_i} \right) \left( \bigotimes_{i<j} CZ_{i,j}^{q_{i,j}} \right).	
\end{equation}
Clearly generators for this group are the $S_i$ gate and the control-$Z$ gate $CZ_{i,j}$, which are well known Cliffords.  Similarly, for $U_F \in \mathcal{D}_3$ we have that 
\begin{equation}
	U_{F} = \left( \bigotimes_{i} T_i^{l_i} \right) \left( \bigotimes_{i<j} CS_{i,j}^{q_{i,j}} \right) \left( \bigotimes_{i<j<k} CCZ_{i,j,k}^{c_{i,j,k}} \right),	
\end{equation}
so as remarked earlier $T_i$, $CS_{i,j}$ and $CCZ_{i,j,k}$ are generators for $\mathcal{D}_3$.  It is widely know that $T$ gates belong to the third level of the Clifford hierarchy, and quick to verify for control-$S$ and control-control-$Z$.  This completes the proof.

It is also an informative exercise to show $\mathcal{D}_3 \subset \mathcal{C}_j$ without a decomposition into generators.  One finds
\begin{equation}
	U_{F}^\dagger X[\vec{m}]	U_{F} X[\vec{m}]   = U_{F'},
\end{equation}
where $X[\vec{m}]:=\otimes_{j=1}^n X_j^{m_j}$ and
\begin{equation}
		F'(\vec{x})=F(\vec{x} \oplus \vec{m}) - F(\vec{x}).
\end{equation} 
Using $\vec{x} \oplus \vec{m} = \vec{x} + \vec{m} - 2  \vec{x}  \wedge  \vec{m}  $ and expanding out $F'(\vec{x})$ into an explicit polynomial, one finds that terms of degree $r$ in the $\vec{x}$ variables carry a prefactor that is a multiple of $2^{r}$.  Therefore, $F'$ can be divided by 2 and remain a weighted polynomial, and so $U_{F'}$ is Clifford.

\section{Evaluating the $P$ matrix}
\label{Pmat}

Here we prove Eq.~(\ref{PmatEq}).  The same proof techniques are used as throughout Sec.~\ref{Sec_construct}, but it is presented here to avoid repetition in the main text. Begin by observing that $P$ can be broken up into four submatrices so that
\begin{equation}
\left( \begin{array}{c}
	\tilde{K}_1 \\
	\tilde{S}_1
\end{array} \right) = \left(  \begin{array}{cc}
	C & 0 \\
	Z & Z \\  	
  \end{array}
\right) ,
\end{equation}
where $C$ is the $k$-by-$4$ matrix
\begin{equation}
	C = (\vec{c}, \vec{c}, \vec{c},\vec{c} ) = \left( \begin{array}{cccc}
 c_1 & c_1 & c_1 & c_1 \\
 c_2 & c_2 & c_2 & c_2 \\
 \vdots & \vdots & \vdots & \vdots \\
 c_k & c_k & c_k & c_k \\
 \end{array}
\right)
\end{equation}
for column vector $\vec{c}=(c_1, c_2, \ldots c_k)^T$, and $Z$ is
\begin{equation}
	Z = \left( \begin{array}{cccc}
 1 & 0 & 0 & 1 \\
 0 & 1 & 0 & 1 \\
 1 & 1 & 1 & 1 
 \end{array}
\right)	.
\end{equation}
Therefore,
\begin{equation}
	| \tilde{K}_1^T \vec{x} \oplus \tilde{S}_1^T\vec{y} | = 	|(C^T\vec{x})\oplus(Z^T \vec{y})| + |Z^T \vec{y}|.
\end{equation}
Using the modular identity, we expand out the first term\begin{equation}
	| \tilde{K}_1^T \vec{x} \oplus \tilde{S}_1^T\vec{y}  | = 	|C^T\vec{x}|-2 |(C^T\vec{x}) \wedge (Z^T \vec{y})| + 2|Z^T \vec{y}|.
\end{equation}
The term $|Z^T\vec{y}|$ will produce some weighted polynomial, and with the added factor 2 this becomes a trivial Clifford term $2|Z^T\vec{y}| \cliff 0$.  The first term equals $|C^T\vec{x}|=4 (\vec{c}^T\vec{x}) $. The factor $(\vec{c}^T\vec{x})$  also corresponds to some weighted polynomial and so with the prefactor 4, it is Clifford.  With these simplifications,
\begin{align}
	| \tilde{K}_1^T \vec{x} \oplus \tilde{S}_1^T\vec{y}  | & \cliff 	-2 |(C^T\vec{x}) \wedge (Z^T \vec{y})|  \\
 & \cliff 	2 |(C^T\vec{x}) \wedge (Z^T \vec{y})| .
\end{align}
Inside the wedge we have $Z^T \vec{y}$, which is explicitly
\begin{align}
		Z^T \vec{y} & = \left( \begin{array}{ccc}
1 & 0 & 1 \\
0 & 1 & 1 \\
0 & 0 & 1 \\
1 & 1 & 1 
 \end{array}
\right) \left( \begin{array}{c}  y_1 \\ y_2 \\ y_3	 \end{array} \right)
= \left( \begin{array}{c}  y_1 \oplus y_3 \\ y_2 \oplus y_3 \\ y_3 
\\ y_1 \oplus y_2 \oplus y_3 \end{array} \right).
\end{align}
The other factor of the wedge is
\begin{equation}
(C^T\vec{x}) =  \left( \begin{array}{c}  \vec{c}^T\vec{x} \\ \vec{c}^T\vec{x} \\ \vec{c}^T\vec{x} \\ \vec{c}^T\vec{x}	 \end{array} \right)
= \left( \begin{array}{c}  \oplus_j c_j x_j \\ \oplus_j c_j x_j \\ \oplus_j c_j x_j \\ \oplus_j c_j x_j	 \end{array} \right).
\end{equation}
Taking the wedge we have
\begin{equation}
	(C^T\vec{x}) \wedge (Z^T \vec{y}) = 	\left( \begin{array}{c}  \vec{c}^T\vec{x} \cdot (y_1 \oplus y_3) \\ \vec{c}^T\vec{x} \cdot (y_2 \oplus y_3) \\ \vec{c}^T\vec{x} \cdot (y_3) 
\\ \vec{c}^T\vec{x} \cdot ( y_1 \oplus y_2 \oplus y_3) \end{array} \right),
\end{equation}
and so 
\begin{align}
	| \tilde{K}_1^T \vec{x} \oplus \tilde{S}_1^T\vec{y}  | & \cliff 2|(C^T\vec{x}) \wedge (Z^T \vec{y})| & =2 	\vec{c}^T\vec{x} \cdot f(\vec{y}),
\end{align}
where
\begin{align}  \nonumber
	f(\vec{y})  =&   (y_1 \oplus y_3) + (y_2 \oplus y_3) + y_3 +
( y_1 \oplus y_2 \oplus y_3)	 \\ \nonumber
 =&  (y_1+y_3-2y_1 y_3)+(y_2 + y_3 - 2y_2 y_3)+ y_3 \\ \nonumber
&+ \big( y_1+y_2+y_3 - 2y_1y_2 - 2y_1 y_3 - 2y_2 y_3  \\  \nonumber
 & + 4 y_1 y_2 y_3 \big) .
\end{align}
Above we have converted from modular math to standard, retaining the brackets to show where terms came from.  Now collecting terms, and including the factor 2 we have
\begin{align}  \nonumber
	2f(\vec{y}) & = 4(y_1+ y_2-y_1y_2)+ 8(y_3 -  y_1 y_3 - y_2 y_3  + y_1 y_2 y_3) \\ \nonumber
 &	= 4(y_1+ y_2-y_1y_2) , 
\end{align}
where in the last line we use that these functions are always taken modulo 8. We must multiply this by $\oplus_i c_i x_i$, which itself expands out to $\sum_i c_i x_i - 2 \sum_{i<j}c_i c_j x_i x_j + \ldots $.  But since everything is modulo 8 only the terms linear in $\vec{x}$ will remain in the following expression
\begin{equation}  \nonumber
	| \tilde{K}_1^T \vec{x} \oplus \tilde{S}_1^T\vec{y}  | \cliff 2 (\vec{c}^T\vec{x}) f(\vec{y})  = 4 \sum_i c_i x_i y_1 y_2,
\end{equation}
which concludes the proof.

\section{Converse proof for controlled-unitaries}
\label{converse}

To prove optimality, we begin by noting that any gate-synthesis matrix $A$ has the form
\begin{equation}
	A_{\mathrm{general}}=\left( \begin{array}{c c c}
B_1 & B_2 & 0 \\
\vec{1}^T & \vec{0}^T 	& l'
 \end{array}
 \right)	,
\end{equation}
for some $B_1$, $B_2$ and $l'$.  We again take $|A_{\mathrm{general}}^T \vec{x}| = |(B_1^T \vec{x}')\oplus(\vec{1}x_k)|+ |B_2^T\vec{x'}| + lx_k$, and simplify it to 
\begin{equation}
  	|A_{\mathrm{general}}^T \vec{x}| \cliff |B_1^T \vec{x}'|+|B_2^T\vec{x'}| + x_k(|\vec{1}|+l')-2x_k |B_1^T \vec{x}'| .
\end{equation}
If this is Clifford equivalent to $F(\vec{x}) =  2 x_k  g(\vec{x}')$, then all the following conditions must hold:
\begin{align}
  	g(\vec{x}')  & \pauli  |B_1^T \vec{x}'| ;\\ \nonumber
    0 &  \cliff x_k(|\vec{1}|+l')  ; \\ \nonumber
    0 &  \cliff |B_1^T \vec{x}'|+|B_2^T\vec{x'}| . 
\end{align}
Remember that Lempel's factorisation method gives an optimal solution of the first equation, and so $\mathrm{col}(B_1) \geq \mathrm{col}(B)$ where $B$ is the optimal solution used above.  The second condition demands that $|\vec{1}|+l' = 0 \pmod{2}$.  The last condition can be written as $|B_1^T \vec{x}'| \cliff |B_2^T\vec{x'}|$.   The $\cliff$ relation is finer than $\pauli$, and so we can infer $|B_1^T \vec{x}'| \pauli |B_2^T\vec{x'}|$.  Therefore, $B_2$ obeys $\mathrm{col}(B_2) \geq \mathrm{col}(B)$ otherwise we would have a contradiction to $B$ being an  optimal solution of $g(\vec{x}') \pauli |B^T \vec{x}'| $.  Since, neither $B_1$ nor $B_2$ can have fewer columns than $B$, and $l'$ is similarly fixed, we see $A_{\mathrm{general}}$ can not outperform the solution given in Eq.~(\ref{Aopt}).

\section{Relationship to triorthogonality}
\label{App::TriOrtho}

Here we illuminate the relationship between the rows of $G$ matrices and the functions they represent.  This will reveal how the triorthogonality condition of Bravyi and Haah relates to our setting. First we need some new notation. Given a binary matrix $G$, we use $\vec{g}^j$ to denote the $j^{\mathrm{th}}$ column vector of $G^T$, so that
\begin{align}
		\vec{g^j} &=(g^j_1, g^j_2, \ldots , g^j_n  )^T \\ \nonumber
		&=(G_{j,1},G_{j,2}, \ldots , G_{j,n})^T .
\end{align}
In other words, $[\vec{g^j}]^T$ is the $j^{\mathrm{th}}$ row vector of $G$.  Again using $| \ldots |$ for the weight of a vector, we have
\begin{equation}
	 | \vec{g^j} |=\sum_{a} g^j_a.
\end{equation}
The symbol $\wedge$ continues to denote element-wise products, so that the $i^{\mathrm{th}}$ element of a wedge is
\begin{equation}
	  [\vec{g^j} \wedge \vec{g^k}]_{i} =  g^j_i g^k_i ,
\end{equation}
which generalises for an arbitrary number of vectors, e.g.
\begin{equation}
	 [ \vec{g^j} \wedge \vec{g^k} \wedge \vec{g^l} ]_i  =  g^j_i g^k_i g^l_i .
\end{equation}
We will show that these wedge products give the coefficients in a weighted polynomial related to $G$.   We begin with
\begin{equation}
  | G^T \vec{z} | = |  \bigoplus_j \vec{g^j} z_j  | = \sum_{i} \left[ \sum_j g^j_i z_j \pmod{2} \right]  ,
\end{equation}
where modulo 2 is only within the brackets.  Typically,  $G$ will be partitioned into $\vec{x}$ and $\vec{y}$.  For now it is easier to ignore the partition.  We proceed by converting the $\pmod{2}$ arithmetic into linear arithmetic.   For three bits this conversion gives that
\begin{equation}
     a_1 \oplus a_2 \oplus a_3 = a_1+a_2+a_3 - 2(a_1 a_2+a_2a_3+a_1 a_3) + 4 a_1 a_2 a_3.
\end{equation}
and more generally
\begin{equation}
     \bigoplus_i a_i = \sum_i a_i - 2 \sum_{i<j}a_i a_j  + 4 \sum_{i<j<k} a_i a_j a_k +\ldots,
\end{equation}
where the dots indicate that there are higher order terms, but these carry prefactors that are multiples of 8 and so will not be relevant here.  Applying this to $   |G^T \vec{z} | \pmod{8}$ we have that
\begin{align}
    | \bigoplus_j \vec{g^j} z_j | &= \sum_j | \vec{g^j} | z_j  - 2 \sum_{i<j} | \vec{g^i} \wedge \vec{g^j} | z_i z_j \\ \nonumber
    &+ 4  \sum_{i<j<h} | \vec{g^i} \wedge \vec{g^j} \wedge \vec{g^h} | z_i z_j z_h  \pmod{8}. 
\end{align}
We see this is a weighted polynomial with coefficients  
\begin{align}
    l_i &:= | \vec{g^i} |  \pmod{8}, \\
    q_{i,j} &:= - | \vec{g^i} \wedge \vec{g^j} |  \pmod{4}, \\
    c_{i,j,k} &:= | \vec{g^i} \wedge \vec{g^j} \wedge \vec{g^k} | \pmod{2} .
\end{align} 
Therefore, $|G^T(\vec{z})|=F(\vec{z})$ where the function $F$ has coefficients determined by considering the row weights, and the weights of wedge pairs and triples.

Next, we translate $F$-quasitransversality into this language
\begin{lem}
\label{Parition_lem}
Let $G$ be a full $\mathbb{Z}_2$-rank matrix with $n$ columns and $r$ rows that is partitioned into $K$ and $S$ so that $G=(\frac{K}{S})$ and $|G^T(\vec{x},\vec{y})|=F(\vec{x},\vec{y}) $ as argued above.  If for all $i$ in the interval $\mathrm{row}[K] < i \leq \mathrm{row}[G]$, and all $j, k$ we have
\begin{align}
    | \vec{g^i} | & = 0 \pmod{2},  \\
    | \vec{g^i} \wedge \vec{g^j} | & = 0 \pmod{2} ,\\
    | \vec{g^i} \wedge \vec{g^j} \wedge \vec{g^k} | & = 0 \pmod{2},
\end{align}
It follows that $F(\vec{x},\vec{y}) \cliff F(\vec{x},\vec{0})$ and so the code is quasitransversal with respect to $F(\vec{x},\vec{0})$.
\end{lem}
The first condition requires that every row in $S$ has even weight.  The second condition tells us that every row in $S$ must have even overlap with every row in the whole matrix $G$. The third condition is that every triple overlap, involving at least one row from $S$, also has even weight.  These conditions are reminiscent of the triorthogonality conditions introduced by Bravyi and Haah.  Indeed, satisfying these conditions is necessary for $G$ to be triorthogonal in their sense, but triorthogonality also requires the second and third condition to extend to all possible pairs and triples of rows (even when $i \leq \mathrm{row}[K]$).  In our more general framework, we allow $G$ to have odd weight overlap of pairs and triples of rows solely within $K$, which results in $CT^{\otimes n}$ implementing multiqubit logical unitaries as we have seen.

Let us now prove the lemma. Recall that the matrix partition of $G$ also splits $\vec{z}$ into $\vec{x}$ and $\vec{y})$. The first condition $| \vec{g^i} |  = 0 \pmod{2}$ for all $\mathrm{row}[K] < i \leq \mathrm{row}[G]$, holds if and only if $L(\vec{z})=L(\vec{x},\vec{y})$ is even valued whenever $\vec{x} \neq 0$. Therefore, $L(\vec{x},\vec{y})=L(\vec{x})+ 2 \tilde{L}(\vec{x},\vec{y})$ for some $\tilde{L}$.  Similarly, the second and third condition are equivalent to the functions $Q$ and $C$ being even whenever $\vec{x} \neq 0$.  Therefore, if the conditions hold then $F$ has the form $F(\vec{x},\vec{y})=F(\vec{x})+ 2 \tilde{F}(\vec{x},\vec{y})$.  This completes the proof.  

We see that our presentation of Thm.~\ref{thm_main} could be stated without reference to phase polynomials and instead in the language of the weight of rows and their overlaps.  Difference audiences may have preferences over the order of presentation, and part of our goal here is to provide a lexicon encompassing these two formalisms.

% \bibliography{MagicLib4}


\end{document}